\documentclass[12pt]{article}
\usepackage{color}
\usepackage{graphicx}
\usepackage{subfigure}
\usepackage{ulem}
\usepackage{amsfonts}
\usepackage{graphicx}
\usepackage{comment}
\newcommand{\be}{\begin{equation}}
\newcommand{\ee}{\end{equation}}
\newcommand{\ba}{\begin{scriptsize} \begin{eqnarray}}
\newcommand{\ea}{\end{eqnarray} \end{scriptsize}}

\usepackage{a4}
\usepackage{cite}
\usepackage{color}
\begin{document}
\title{Thermodynamic Geometry and Topological Einstein-Yang-Mills Black Holes}
\author{}
\date{Stefano Bellucci$^{}$\thanks{\noindent bellucci.stefano@lnf.infn.it}\
\ and Bhupendra Nath Tiwari$^{}$\thanks{\noindent bntiwari.iitk@gmail.com}\\
\vspace{0.50 cm} \vspace{0.20 cm}
$^{}$INFN-Laboratori Nazionali di Frascati\\
Via E. Fermi 40, 00044 Frascati, Italy.\\ \vspace{0.5cm}
}
\vspace{0.10 cm}
\maketitle \abstract{From the perspective of the statistical
fluctuation theory, we explore the role of the thermodynamic
geometries and vacuum (in)stability properties for the topological
Einstein-Yang-Mills black holes. In this paper, from the
perspective of the state-space surface and chemical Weinhold
surface, we provide the criteria for the local and global
statistical stability of an ensemble of topological
Einstein-Yang-Mills black holes in arbitrary spacetime dimensions
$D\ge 5$. Finally, as per the formulations of the thermodynamic
geometry, we offer a parametric account of the statistical
consequences in both the local and global fluctuation regimes of
the topological Einstein-Yang-Mills black holes.}

\vspace{1.5cm} \textbf{Keywords:} \textit{Thermodynamic Geometry;
Topological Einstein-Yang-Mills Black Holes; Higher Dimensional Gravity;
Cosmological Constant.} \\

\section{Introduction}

Thermodynamic geometry \cite{RuppeinerRMP,RuppeinerPRD78,RuppeinerA20,RuppeinerPRL,
RuppeinerA27,RuppeinerA41,bntuns,bnt,0606084v1,SST,BNTBull,BNTBull08,BNTBullcorr,
Hawking,more1,more2,grqc0601119v1,grqc0512035v1,grqc0304015v1,0510139v3,bntiit,
Weinhold1,Weinhold2,BNTSBVC,talk,Quark,EEEE1,EEEE2} plays important role in understanding
the stability and phase structure properties of black holes. There have been several
investigations made in this direction, which explore the thermodynamic structures of
black holes in gerenal relativity, string theory and M-theory. In this paper,
we examine thermodynamic structures of topological Einstein-Yang-Mills black holes.

From the perspective of the  $SU(2)$ gauge theory, we explore the thermodynamic
properties of the black hole solutions in non-abelian gauge theory. In particular,
the present consideration investigates the thermodynamic stability structures of a
class of two parameter extremal black hole configurations which carry an electric
charge $e$ and the cosmological constant $\Lambda:= kn(n-1)/l^2$. Here, for a given
$(n+ 1)$-dimensional topological Einstein-Yang-Mills black hole, the symbol $k$
takes vales over the set $\{-1, 1\}$ respectively for the asymptotically Anti-de-Sitter
(AdS) and de-Sitter (dS) solutions. From the perspective of the  $SU(2)$ gauge theory,
Ref. \cite{1} offers the subject matter, namely, the solitonic features towards such a
consideration. In this concern, the work of Yasskin \cite{2} shows the corresponding
colored black hole solutions with $SO(3)$ gauge group.

For a given colored black hole solution, in the above setup, the black hole uniqueness theorem leads
to the fact that arbitrary dimensional topological solutions of the Einstein-Yang-Mills black
holes are hairy in their character. In contrast to the Kerr-Newman family, the exterior geometry
of the both the above family of solutions, apart from the global asymptotic charges, require
additional parameters of the metric tensor and the corresponding matter fields \cite{8,9,10}.
The nature of these solutions have been extended for the higher derivative corrections.
Namely, Refs. \cite{16,17} provide the contribution arising from the Gauss-Bonnet terms.
Refs. \cite{11,12,13,14,15} describe the corresponding cosmological
constant contributions to the entropy and ADM mass of the topological Einstein-Yang-Mills
black holes black holes. It is worth mentioning further that all the aobve solutions indicate a
non-trivial domain of unstability for the non-negative cosmological constant solutions,
see for a reference \cite{18}. Refs. \cite{14,19} provide the (un)stability structures
of the associtated negative cosmological constant black holes.

From the perspective of gauge field theories, for a given N-parameter semi-simple Lie gauge group,
the action of $(n+1)$- dimensional Einstein-Yang-Mills gravity theory discribes black hole solutions.
This consideration involves three type of contributions to the Lagrangian of the theory. These terms
are respectively the contributions arising from the Ricci scalar, cosmological constant (or $AdS$ curvature)
and the Yang-Mills field strength tensor. The equations of the motion of such a Langragian are obtained via
variations of the background space-time metric tensor and the underlying Yang-Mills gauge fields.
For given gauge fields and space-time metric tensor, the equations of motion render as a set of coupled
Yang-Mills equations with sources and the Einstein field equations with cosmological constant.
Such a theory can be simplified via the consideration of the Cartan’s criteria. Namely, this yields to the
fact that both the energy momentum tensor and current of the theory can be expressed in terms of the structure
constants of the gauge group, gauge coupling constants and underlying gauge fields. From the consideration
of Wu-Yang ansatz \cite{30}, Ref. \cite{BD} provides explicit topological character of Einstein-Yang-Mills
black holes and thus one obtains black hole solutions in four, five and arbitrary spacetime dimensions.
It is worth mentioning that the determination of the currents and the gauge fields of the theory requires
specification of structure constants of the gauge group. Moreover, the static space-time metric tensor of
such a black hole, which possesses the hyperbolic horizon, arises by solving the corresponding cosmological
Einstein field equations for specific components of the background space-time metric tensor.
In this perspective, Ref. \cite{31} provides a class of static non-abelian five-dimensional
black holes in the theory of $\mathcal N= 8$ maximal gauged supergravity.

Notice further that the Ref. \cite{BD} provides an appropriate plateform to study the
thermodynamic geometric properties of the non-abelian black holes. In any case, as a generalization
of the abelian Einstein-Maxwell gravity, these black holes are based on the analysis of the
Einstein field equations. For nonlinear field sources with generic topological nature,
we shall offer exact thermodynamical properties of the standard four and higher
dimensional black holes in the theory of the Yang-Mills gravity. Specifically, for given
$(n + 1)$-dimensional topological Einstein-Yang-Mills black holes with a negative cosmological
constant, the thermodynamics lies on the properties of the Einstein field equations,
gauge current and stress–energy tensor for arbitrary finite semi-simple gauge group.
In fact, the consideration of Ref. \cite{BD} leads to explicit horizon properties of the
four dimensional solution with $SO(2,1)$ gauge symmetry. The above research direction
further continues towards the topological properties of a family of logarithmically
corrected black hole solutions in the five and other higher dimensional generalizations.
The examination of the five-dimensional static black holes provides interesting issues
for the hyperbolic horizon spherically symmetric solutions with $SO(3,1)$ gauge isometries.
In this perspective, we refere Ref. \cite{BD} for explicit expressions of the mass, entropy
and Hawking temperature of the black holes of the present interest. Namely, the consideration
of the thermodynamic geometry offers further issues for the non-static asymptotically de Sitter
solutions in six and higher spacetime dimensional Einstein-Yang-Mills gravity with a non-negative mass,
which we relagate to the section $3$ and $4$ of this paper. The motivation of the present paper lies
in the fact that these solutions describe black holes if the corresponding solutions exist in the
Einstein-Maxwell gravity \cite{BD}. Namely, it is worth mentioning that the notion of the thermodynamic
geometry opens new direction to examine the stability properties of topological Einstein-Yang-Mills black holes
in arbitrary $(n+1)$-dimensional spacetime with $SO(n(n-1)/2-1, 1)$ semi-simple gauge group symmetries.

Here, our goal is to analyze the statistical nature of the
topological Einstein-Yang-Mills black holes, in general. Namely,
we wish to explicate the local and global statistical stability of
an ensemble of such black holes in arbitrary spacetime dimensions
$D \ge 5$. Our framework allows one to geometrically explore the
nature of the local and global statistical correlations about a
fixed vacuum of the nonabelian Yang-Mills gauge theory containing
Einstein-Yang-Mills black holes. As per the quantitative analysis
of the thermodynamic geometric model provided in the section $2$,
this paper offers an explicit realization of the statistical
(in)stabilities. For the given black hole entropy and mass, we
shall illustrate that the parametric fluctuations are intrinsic
geometric in the nature. Subsequently, the framework of
fluctutation theory is capable of offering an appropriate account
of the statistical properties of all finite dimensional
topological Einstein-Yang-Mills black hole configurations. From
the perspective of the thermodynamic geometries and fluctuation
theory, the statistical ensmble (in)stabilities via the Ruppeiner
geometry and the corresponding Legendre transformed Weinhold
geometry of the arbitrary finite spacetime dimensional topological
Einstein-Yang-Mills black hole configurations are of our
particular interest in the consideration of the present paper.

The rest of the paper is organized as follow. In section $2$, we
provide a brief review of the fluctuation theory of the two
parameter black hole configuration and thereby specialize it from
the perspective application of the thermodynamic Riemannian
geometries. In section $3$, we analyze the Ruppeiner geometry for
the five dimensional topological Einstein-Yang-Mills black hole
configuration and thereby extend our consideration for the
ensemble of arbitrary finite dimensional topological
Einstein-Yang-Mills black holes. In section $4$, we explore the
above consideration from the perspective of the Weinhold geometry,
where we first analyze the ensemble of the five dimensional
topological Einstein-Yang-Mills black holes and subsequently
consider the case of arbitrary finite dimensional topological
Einstein-Yang-Mills black hole configuration. Finally, in section
$5$, we present our conclusions and some remarks for a future
investigation.

\section{Thermodynamic Geometry}

In what follows next that we consider an intrinsic Riemannian geometric
model  whose covariant  metric tensor  may be  defined as  the Hessian
matrix  of  the  entropy  with  respect  to  an  arbitrary  number  of
parameters  characterizing a black hole system  of the statistical interest.
We shall examine the nature of the fluctuation properties with a minimal number of parameters,
such as fixed volume system, characterizing the thermodynamics of the associated equilibrium
statistical configuration of an ensemble of arbitrary finite dimensional topological
Einstein-Yang-Mills black holes. In particular, let us define the entropy representation
of the chosen configuration with the entropy $ S(x_i)$ for a given set of parameters
of the configuration, \textit{viz.} entropy, temperature, mass and charges as
the set $ \lbrace x_i \rbrace  $. Here, the parameters $x_i$, for a given set of
events $\{i_j\ |\ j \in \Lambda \} \subset Z$ with a finite set $\Lambda$,
are determined by an equilibrium distribution function.

The set $\{x_i \}$, for $i = 1,...,n$, when treated as a set of
extensive thermodynamic variables, forms coordinate charts for the
corresponding intrinsic manifold. In this sense, an appropriate
choice of the parameters $x_i$, characterizes the entropy of the
system. This charecterization of the fluctuations of an ensemble
of finite dimensional topological Einstein-Yang-Mills black holes
renders into the the so called Ruppeiner geometry
\cite{RuppeinerRMP,RuppeinerPRD78,RuppeinerA20,RuppeinerPRL,
RuppeinerA27,RuppeinerA41}. In general, the components of the
covariant Ruppeiner metric tensor are defined as

\begin{eqnarray}
g_{ij}:= -\frac{\partial^2 S(\vec{x})}{\partial x^i \partial x^j},
\end{eqnarray}

where the vector $\vec{x} =(x^i) \in M_n $. Explicitly, for the
case of  the two  dimensional intrinsic Riemannian geometry
parametrized  by $\vec{x} =(x_1,x_2)  \in M_2  $,  the components
of  the thermodynamic Ruppeiner metric tensor are given by

\begin{eqnarray}
\label{eq2}
g_{x_1 x_1}&=& -\frac{\partial^2 S}{\partial x_1^2}, \nonumber\\
g_{x_1 x_2}&=& -\frac{\partial^2 S}{{\partial x_1}{\partial x_2}}, \nonumber\\
g_{x_2 x_2}&=& -\frac{\partial^2 S}{\partial x_2^2}.
\end{eqnarray}

Notice that the components of the intrinsic metric tensor are
associated to the respective pair correlation functions of the
concerned entropy flow. It is worth mentioning that the
co-ordinates of the underlying fluctuations lie on the surface of
the parameters $\{x_1, x_2\}$, which in the statistical sense,
gives the origin of the fluctuations in the vacuum topological
Einstein-Yang-Mills black holes. This is because the components of
the Ruppeiner metric tensor comprise the Gaussian fluctuations of
the degeneracy of the microstates, which is a function of the
parameters of the associated macroscopic black hole configuration.
For a given black hole, the local stability of the underlying
statistical system requires both the principle components to be
positive. In this concern, the diagonal components of the
Ruppeiner metric tensor, $\{ { g_{x_ix_i} \mid i = {1,2}}\}$
signify the heat capacities of the chosen system. From the
perspective of the fluctuation theory, it is required that the
diagonal components of the Ruppeiner metric tensor remain positive
definite quantities, \textit{viz.} we must have

\begin{eqnarray}
g_{x_ix_i} &>& 0, \ i= \ 1,2
\end{eqnarray}

for the existance of the local statistical stability of the two parameter black holes.
In this case, we see that the determinant of the metric tensor can be given as

\begin{eqnarray} \label{determinant}
\Vert g \Vert &= &S_{x_1 x_1}S_{x_2 x_2}- S_{x_1 x_2}^2.
\end{eqnarray}

The Christoffel connections $\Gamma_{ijk}$, Riemann curvature tensors $R_{ijkl}$,
Ricci tensors $R_{ij}$ and the scalar curvature $ R $ of the two dimensional thermodynamic
geometry $(M_2,g)$ can be computed further. With the above notion, we find that
the scalar curvature can be shown to be

\begin{eqnarray}
R&=& \frac{1}{2\Vert g \Vert^2 } \bigg(S_{x_2 x_2}S_{x_1x_1x_1}S_{x_1 x_2 x_2}+ S_{x_1 x_2}S_{x_1x_1x_2}S_{x_1x_2x_2}
\nonumber\\ &&+ S_{x_1x_1}S_{x_1x_1x_2}S_{x_2x_2x_2}-S_{x_1x_2}S_{x_1x_1x_1}S_{x_2x_2x_2}\nonumber\\ &&-
S_{x_1x_1}S_{x_1x_2x_2}^2- S_{x_2x_2}S_{x_1x_1x_2}^2 \bigg).
\end{eqnarray}

Interestingly, the relation between the thermodynamic scalar curvature
and the Riemann curvature  tensor for any two  dimensional intrinsic
Riemannian geometry is given (see for details \cite{0606084v1, bnt}) by

\begin{eqnarray}
 R=\frac{2}{\Vert g \Vert}R_{x_1x_2x_1x_2}.
\end{eqnarray}

It is  worth mentioning that the relationship of a  non-zero scalar curvature
with an  underlying interacting  statistical system remains valid for
higher  dimensional  intrinsic  manifolds, as well. Namely, the connection
of  a divergent  (scalar) curvature with  phase transitions  can be analyzed
from  the  Hessian  matrix  of  the  considered  fluctuating entropy.
In the sense of the state-space  fluctuations, such a consideration of the
statistical fluctuations requires an ensemble of vacuum black hole configurations.
Specifically, the present article divulges the underlying geometric description
in the Gaussian approximation. Such an  analysis is thus concerned in the neighborhood
of the small probability fluctuations. Hereby, the present consideration takes into
account the scales that are larger than the correlation length of the system,
in which a few microstates do not dominate the entire macroscopic phase-space
configuration of the chosen dimensional topological Einstein-Yang-Mills
black hole ensemble.

As per the Gaussian distribution theory, the  thermodynamic
Ruppeiner metric may be expressed as the second moment of the
quadratic fluctuations or the statistical parametric pair
correlation functions. Indeed, an explicit evaluation shows the
components of the inverse metric tensor are

\begin{eqnarray}
g^{ij}=  \langle x^i \vert x^j\rangle,
\end{eqnarray}

where  $\lbrace  x_i  \rbrace  $'s  are  the  extensive
thermodynamic variables conjugate  to the  intensive variables
$\lbrace  X_i \rbrace $. Moreover, such Riemannian structures may
also be expressed in terms of a suitable thermodynamic potential
obtained by a Legendre transformation. In section $4$, we
explicate such a consideration for the Weinhold geometry, arrising
from the fluctuations of the mass of the topological
Einstein-Yang-Mills black hole in spacetime dimensions $D\ge5$.
For a given statistical ensemble, it is worth mentioning that in
the above intrinsic geometric setup corresponds to certain general
coordinate transformations on the space of equilibrium states.

\section{Ruppeiner Geometry}

In this section, we examine thermodynamic fluctuation properties of the
topological Einstein-Yang-Mills black hole as per the prescription
of the thermodynamic Rupppenier geometry. Following the explications
in the previous section, we first consider the analysis of the vacuum
statistical correlation for an ensemble of five dimensional topological
Einstein-Yang-Mills black hole. For the given vacuum parameters,
we shall exhibit that the parametric thermodynamic geometry is well capable
to describe the perspective statistical (in)stability corresponding to the
topological Einstein-Yang-Mills black hole configurations. Subsequently,
we analyze the above properties for an ensemble of arbitrary spacetime
dimensional topological Einstein-Yang-Mills black holes.

\subsection{Five Dimensional Black Holes}

From the Ref. \cite{BD}, the entropy of a topological Einstein-Yang-Mills
black hole in spacetime dimension $D=5$ can be expressed as

\begin{scriptsize} \begin{eqnarray}
\mathrm{S}(l, \,e) := {\displaystyle \frac {1}{32}} \,\mathit{V_3}
\,l^{3}\,(1 + \sqrt{1 + {\displaystyle \frac {8\,e^{2}}{l^{2}}}})^{3}.
\end{eqnarray} \end{scriptsize}

As per the notion of the Gaussian fluctuation theory, we see that
the line elements associated with the Ruppeiner geometry is given
by

\begin{scriptsize} \begin{eqnarray}
\mathit{\ ds}^{2}= - ({\frac {\partial ^{2}}{\partial e^{2}}}\,
\mathrm{S}(l, \,e))\,\mathit{\ d}\,e^{\mathit{2\ }} - 2\,( {\frac
{\partial ^{2}}{\partial l\,\partial e}}\,\mathrm{S}(l, \,
e))\,\mathit{\ d}\,e^{\ }\,\mathit{d\ }\,l^{\ } - ({\frac {
\partial ^{2}}{\partial l^{2}}}\,\mathrm{S}(l, \,e))\,\mathit{\ d
}\,l^{\mathit{2\ }}.
\end{eqnarray} \end{scriptsize}

Before proceeding further, we introduce the following scaling factor

\begin{scriptsize} \begin{eqnarray} \label{factor5}
\mathrm{f}:= \sqrt{{\displaystyle \frac{l^{2} + 8\,e^{2}}{l^{2}}}}.
\end{eqnarray} \end{scriptsize}

With this convention, we find the following components of the
Ruppeiner metric tensor

\begin{scriptsize} \begin{eqnarray}
{\mathit{g\ }_{\mathit{ll}}}&=& - {\displaystyle \frac {3}{8}} \,\mathit{V_3}\,
{\displaystyle \frac {(6\,e^{2}\,\mathrm{f} + 10\,e^{2} + l^{2}\,\mathrm{f} + l^{2})\,l\, (1 +
\mathrm{f})}{\mathrm{f}\,(l^{2} + 8\,e^{2})}},\\ \nonumber
{\mathit{g\ }_{\mathit{el}}}&=& - {\displaystyle \frac {3}{4}} \,\mathit{V_3}
{\displaystyle \frac {(16\,e^{2} + l^{2} + l^{2}\,\mathrm{f})\,e\,(1 +
\mathrm{f})\, }{\mathrm{f}\,(l^{2} + 8\,e^{2})}},\\ \nonumber
{\mathit{g\ }_{\mathit{ee}}}&=& - {\displaystyle \frac {3}{4}} \,\mathit{V_3}
{\displaystyle \frac {l\,(16\,e^{2}\,\mathrm{f} + l^{2} + l^{2}\,\mathrm{f})\,
(1 + \mathrm{f})\,}{\mathrm{f}\,(l^{2}+ 8\,e^{2})}}.
\end{eqnarray} \end{scriptsize}

\begin{figure}
\vspace*{-0.5cm}
\includegraphics[width=15.0cm,angle=-0]{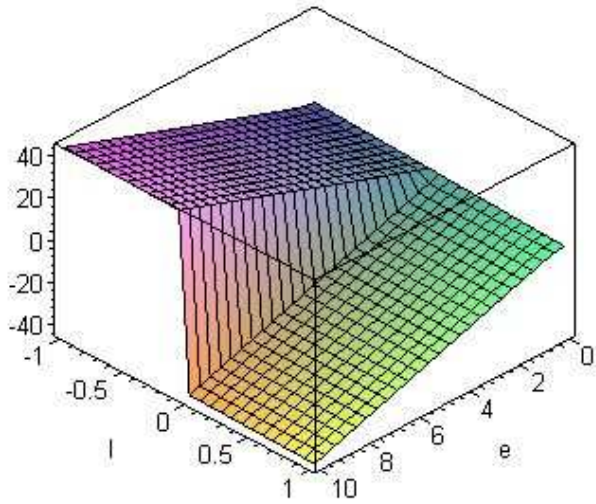}  \vspace*{-5.5cm}
\caption{The $ee$ component of Ruppeiner metric tensor plotted as
the function of $\{ e, l \}$, describing the fluctuations in five
dimensional topological Einstein-Yang-Mills black hole
configurations.} \label{5dree} \vspace*{-0.5cm}
\end{figure}

\begin{figure}
\hspace*{0.5cm}
\includegraphics[width=15.0cm,angle=-0]{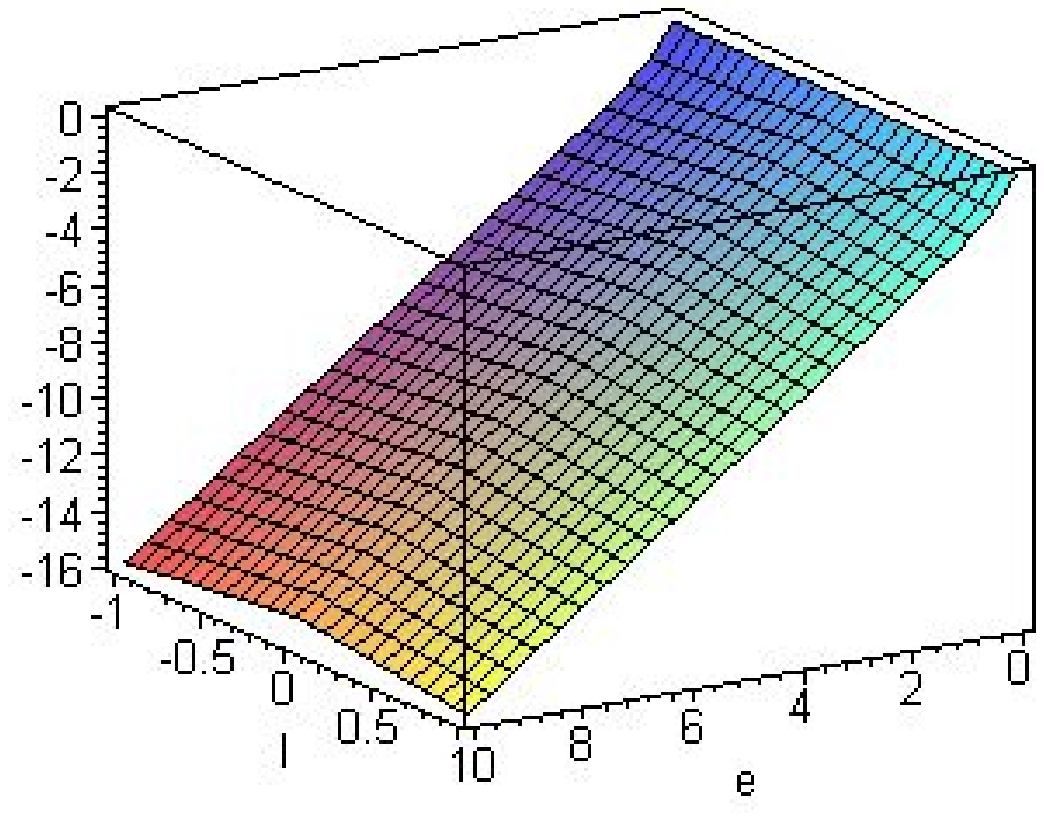}  \vspace*{-5.5cm}
\caption{The $el$ component of Ruppeiner metric tensor plotted as
the function of $\{ e, l \}$, describing the fluctuations in five
dimensional topological Einstein-Yang-Mills black hole
configurations.} \label{5drel} \vspace*{-0.5cm}
\end{figure}

\begin{figure}
\hspace*{0.5cm}
\includegraphics[width=15.0cm,angle=-0]{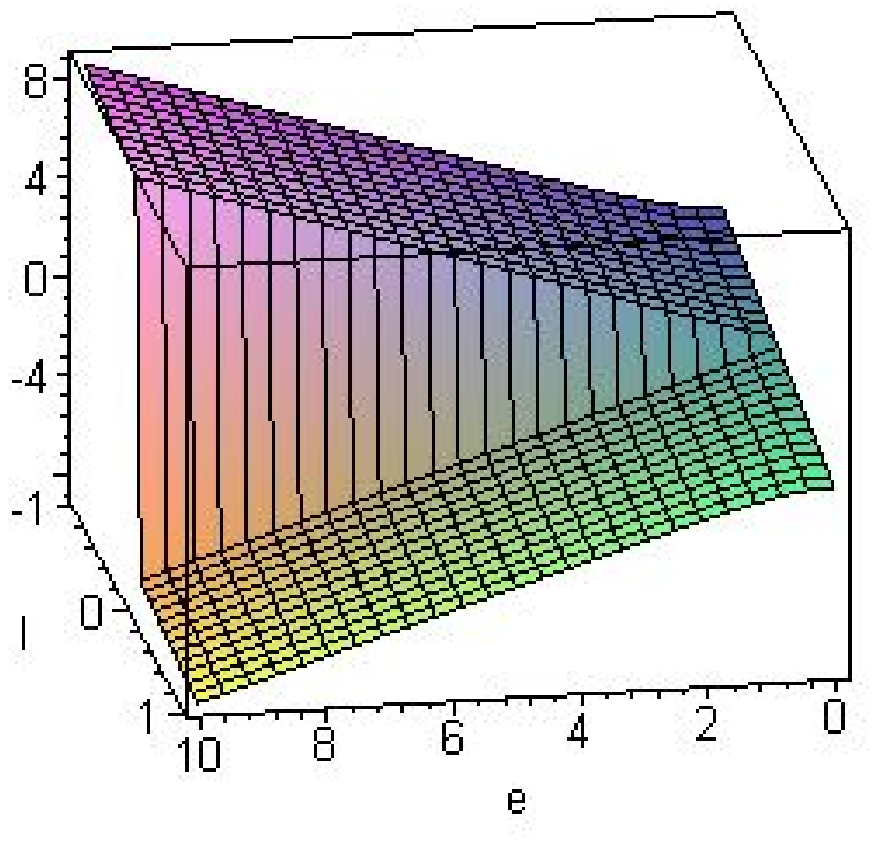}  \vspace*{-5.5cm}
\caption{The $ll$ component of Ruppeiner metric tensor plotted as
the function of $\{ e, l \}$, describing the fluctuations in five
dimensional topological Einstein-Yang-Mills black hole
configurations.} \label{5drll} \vspace*{-0.5cm}
\end{figure}

For a given $\mathit{V_3}$, in order to analyze the instability occurring due to a
entropy fluctuations with respect to the electric charge $e$ and cosmological constant
parameter $l$, the figures Figs.(\ref{5dree}, \ref{5drll}) show the fluctuations
in the diagonal components $\{g_{ee}, g_{ll} \}$ of the metric tensor.
The value of $\mathit{V_3}$ depends on the phase-space of the black hole,
and thus it may vary from vacuum to vacuum, however the procedure of the
state-space analysis remains the same. In the regime of $l \in (-1,1)$ and $e \in (0,10)$,
we notice that the amplitude of $\{g_{ee} \}$ takes a value between
$\{-40, +40\}$. In this range of the parameters $\{e, l\}$, we find that
the mix component $\{ g_{el} \}$ lies in the range of $\{-16, 0 \}$. In this case,
we see that the range of the growth of the amplitude of $\{g_{ll} \}$ remains in the
regime of $\{-8, +8\}$ for the parameters $\{e, l\}$. Explicitly, this signifies that
the five dimensional topological Einstein-Yang-Mills black holes are thermodynamically
unstable in the limit of a large $e$ and a positive $l$. Thus, the order higher entropy
corrections are required for a large $e$ in order to stabilize the five dimensional
topological Einstein-Yang-Mills black hole system, which can easily be extract from
positivity of the the components of the state-space metric tensor. Similarly, the
Fig.(\ref{5drel}) shows the nature of $\{ g_{el}\}$ component of the state-space metric tensor.
We find that the mix component $\{ g_{el}\}$ takes an uniform decreasing value from zero
to $- 16$ in both the limit of the parameter $l$ and for the increasing value of $e$.
In this limit of $\{e, l\}$, the local fluctuation of the entropy of the
five dimensional topological Einstein-Yang-Mills black hole as depicted in the
Figs.(\ref{5dree}, \ref{5drll}, \ref{5drel}) illustrate the state-space stability
properties of the  the five dimensional topological Einstein-Yang-Mills black hole ensemble.
In short, the self pair fluctuations involving $\{e, l\}$, as defined by the metric tensor
$\{g_{ij}\ |\ i, j= e, l \}$, have both the positive and negative numerical values,
and thus the five dimensional topological Einstein-Yang-Mills black hole
are stable only in a particular domain of the vacuum parameters.

\begin{figure}
\hspace*{0.5cm}
\includegraphics[width=15.0cm,angle=-0]{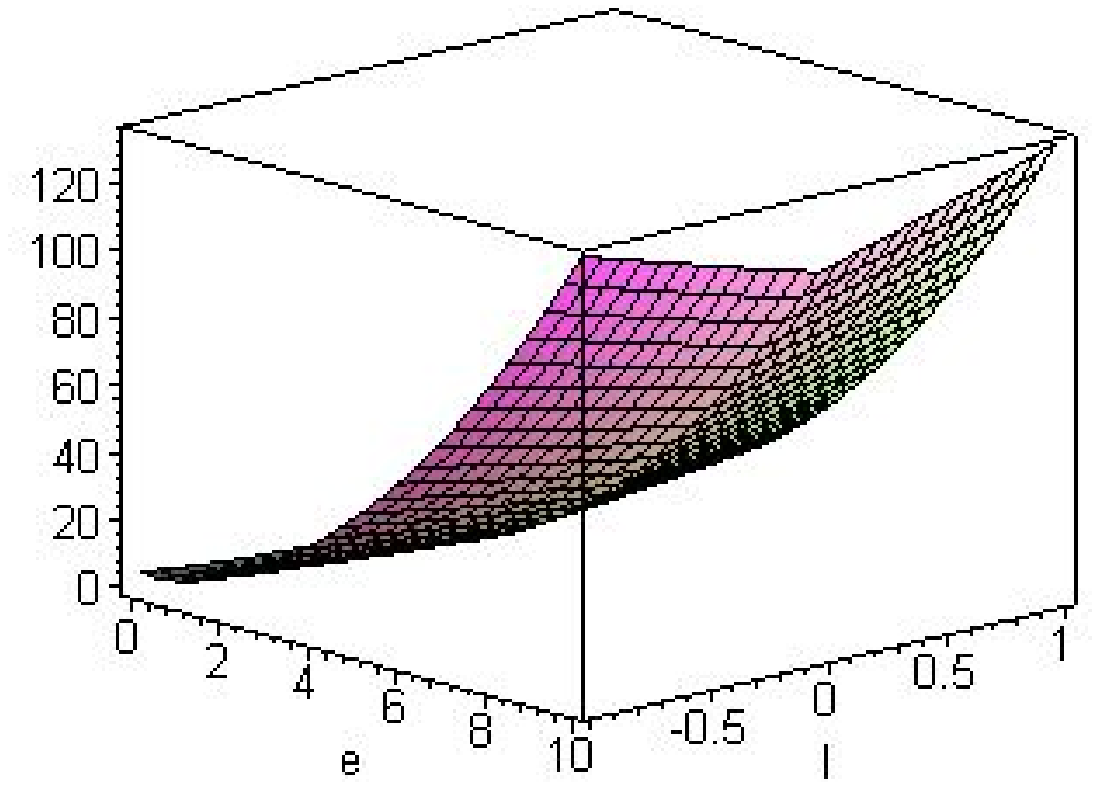}  \vspace*{-5.5cm}
\caption{The determinant of Ruppeiner metric tensor plotted as the
function of $\{ e, l \}$, describing the fluctuations in five
dimensional topological Einstein-Yang-Mills black hole
configurations.} \label{5drdetg} \vspace*{-0.5cm}
\end{figure}

From the above expression of the metric tensor, it is not difficult to see that
the determinant of the metric tensor can be expressed in the following form

\begin{scriptsize} \begin{eqnarray}
\mathit{g\ }={\displaystyle \frac {9}{16}}\,{\displaystyle \mathit{V_3}^{2}\,
\frac{(1 + \mathrm{f})^{2}}{\mathrm{f}^{3}\,(l^{2} + 8\,e^{2})}
\bigg(( 16\,\mathrm{f} +38)\,e^{4} + (10\,\mathrm{f}+ 14)\,l^{2}\,e^{2}
+ (\mathrm{f} +1)\,l^{4}\bigg) },
\end{eqnarray} \end{scriptsize}

where the factor $\mathit{f}$ is defined as per the Eqn. (\ref{factor5}).
Herewith, we see that the five dimensional topological Einstein-Yang-Mills
black holes remain stable under the effect of underlying thermodynamical
fluctuation of $\mathit{\{e, l\}}$ in a chosen domain.

In this case, the ensemble stability of the five dimensional topological
Einstein-Yang-Mills black holes can be determined in terms of the values
of the regulation parameters $e, l$. This follows from the behavior of
the determinant of metric tensor. Notice that the determinant of the
metric tensor tends to a well-defined positive value when the vacuum parameters
take relatively larger absolute values, \textit{viz.} $e \rightarrow 10 $ and
$l \rightarrow \pm 1$. For $e \in (0, 10)$ and $l \in (-1, 1)$, the Fig.(\ref{5drdetg})
shows that the determinant of the metric tensor lies in the interval $(0, 140)$.
In fact, we find that the positivity of $g$ increases as the values of $(e, l)$ are increased
from origin to $(10, \pm 1)$. In such cases, we find that the surface defined by the
fluctuations of $\{e, l\}$ is stable. When only one of the parameter is allowed
to vary, the stability of the underlying black hole configuration is determined by
the positivity of the first principle minor. In other words, this amounts to the
positivity statement of the $ee$ component of the Ruppnier metric tensor.
The above graphical properties and positivity of the state-space quantities
provide the underlying stability properties of the two parameter topological
Einstein-Yang-Mills black holes in five spacetime dimensions.

To examine the metric structure properties of the above fluctuations,
we may compute the fluctuations of the metric tensor $\mathit{g_{ij}}$.
For a given intrinsic surface of $\mathit{\{e, l\}}$, these fluctuations
are precisely given by the following Christofel symbols

\begin{scriptsize} \begin{eqnarray}
{\Gamma _{\mathit{eee}}}&=& - 24\,\mathit{V_3}\,{\displaystyle \frac {(16\,e^{2}
 + 5\,l^{2})\,e^{3}\,}{\mathrm{f}\,(l^{2} + 8\,e^{2})^{2}\,l}},\\ \nonumber
{\Gamma _{\mathit{eel}}}&=& - {\displaystyle \frac {3}{4}} \,\mathit{V_3}\,
{\displaystyle \frac {(64\,\mathrm{f}\,e^{4} + 20\,l^{2}\,e^{2} + 16\,l^{2}\,
\mathrm{f}\,e^{2} + l^{4}\,\mathrm{f} + l^{4}
)\,}{\mathrm{f}\,(l^{2} + 8\,e^{2})^{2}}},\\ \nonumber
{\Gamma _{\mathit{ele}}}&=& - {\displaystyle \frac {3}{4}} \,\mathit{V_3}\,
{\displaystyle \frac {(64\,\mathrm{f}\,e^{4} + 20\,l^{2}\,e^{2} + 16\,l^{2}\,
\mathrm{f}\,e^{2} + l^{4}\,\mathrm{f} + l^{4})\,}{\mathrm{f}\,(l^{2}
+ 8\,e^{2})^{2}}},\\ \nonumber
{\Gamma _{\mathit{ell}}}&=& - 72\,\mathit{V_3}\, {\displaystyle \frac {e^{5}\,
}{\mathrm{f}\,(l^{2} + 8\,e^{2})^{2}\,l}},\\ \nonumber
{\Gamma _{\mathit{lle}}}&=& - 72\,\mathit{V_3}\, {\displaystyle \frac {e^{5}\,
}{\mathrm{f}\,(l^{2} + 8\,e^{2})^{2}\,l}},\\ \nonumber
{\Gamma _{\mathit{lll}}}&=& - {\displaystyle \frac {3}{8}}\,\mathit{V_3} \,
{\displaystyle \frac {(120\,e^{4} + 64\, \mathrm{f}\,e^{4} + 20\,l^{2}\,e^{2} + 16
\,l^{2}\,\mathrm{f}\,e^{2} + l^{4}\,\mathrm{f} + l^{4})}{\mathrm{f}\,(l^{2} + 8\,e^{2})^{2}}}.
\end{eqnarray} \end{scriptsize}

In this case, we find that the underlying thermodynamic
configuration has no global fluctuation and the associated
correlation length vanishes identically with the following
Ruppeiner scalar curvature

\begin{scriptsize} \begin{eqnarray}
\mathit{R\ }(e, l)=0, \ \forall \ (e,l)\ \in \ \mathcal{M}_2.
\end{eqnarray} \end{scriptsize}

The global stability properties of the two parameter topological
Einstein-Yang-Mills black holes in five spacetime dimensions
follow from the underlying state-space scalar curvature. As, in
this case, we find that the scalar curvature vanishes identically
for all values of the black hole parameters. This shows that the
fluctuating five dimensional topological Einstein-Yang-Mills black holes
correspond to a noninteracting statistical configuration. In short, the above
observations of the state-space geometry indicates that the five dimensional
topological Einstein-Yang-Mills black holes are although noninteracting
in the global sense, however they correspond to a stable statistical configuration
in a specific domain of the vacuum parameters. Namely, when the parameter $\{e, l \}$
are allowed to fluctuate, we see that there exists certain domain of the vacuum parameters
in which some of the component can fail to remain positive and a local statistical
instability. This observation follows from the fact that there are non-trivial instabilities
at the local level of the vacuum fluctuations. From the above observation,
it can be seen that the iterative procedure of vacuum parameters can be replaced
by a statistically directed method of the state-space geometry. This formulation
incorporates fluctutation of the parameters which follows the non-linearity Gaussian
approximation about an equilibrium topological Einstein-Yang-Mills black hole system.

Without loss of the generality, for the pictorial depictationa of the thermodynamic quantities,
we may as well choose the phase space volume to be unity, \textit{viz.}, $\mathit{V_3} := 1$.

\subsection{Higher Dimensional Black Holes}
In this subsection, we illustrate the role of state-space geometry
to the arbitrary higher dimensional topological
Einstein-Yang-Mills black holes. In the highly growing spacetime
dimension, the entropy maximisation is necessary in order to
define the statistically stable limit of the field theory vacuum.
Notice that, for the entropy maximization procedure, the Ruppeiner
geometric state-space constraints as defined in the section $2$
are governed by the entropy flow equations. From the Ref.
\cite{BD}, the entropy of a topological Einstein-Yang-Mills black
hole in any spacetime dimension $D=1+n$ can be expressed as

\begin{scriptsize} \begin{eqnarray} \label{entropyn}
\mathrm{S}(l, \,e) := {\displaystyle \frac {1}{8}} \,V\,\sqrt{2}
\,\sqrt{{\displaystyle \frac {n - 2}{n}} }\,l\,(1 + \sqrt{1 +
{\displaystyle \frac {4\,n\,e^{2}}{(n - 2)\,l^{2}}} })^{(n - 1)}.
\end{eqnarray} \end{scriptsize}

To simplify the subsequest expression, we define a level function $f_n$ as

\begin{scriptsize} \begin{eqnarray}
\mathrm{f_n}:= (l^{2} + 4\,e^{2})\,n - 2\,l^{2}.
\end{eqnarray} \end{scriptsize}

Thence, from the definition of Hessian of the entropy Eqn. (\ref{entropyn}),
we find the following expressions for the componets of the metric tensor

\begin{scriptsize} \begin{eqnarray} \label{metricrupn}
{\mathit{g\ }_{\mathit{ll}}}&=& - {\displaystyle \frac {1}{2}}
\sqrt{2}\,(4\,n\,e^{2}\,\sqrt{{\displaystyle \frac {\mathrm{f_n}
}{(n - 2)\,l^{2}}} } + l^{2} + l^{2}\,\sqrt{{\displaystyle
\frac {\mathrm{f_n}}{(n - 2)\,l^{2}}} })\,V\,\sqrt{
{\displaystyle \frac {n - 2}{n}} } \nonumber \\ &&
(1 + \sqrt{{\displaystyle \frac {\mathrm{f_n}}{(n - 2)\,l^{2}}} }
)^{(n - 3)}\,(n - 1)\,n\,e^{2} \left/ {\vrule
height0.84em width0em depth0.84em} \right. \!  \! (\sqrt{
{\displaystyle \frac {\mathrm{f_n}}{(n - 2)\,l^{2}}} }\,l^{3}\,
\mathrm{f_n}), \nonumber \\
{\mathit{g\ }_{\mathit{el}}}&=& {\displaystyle \frac {1}{2}} \sqrt{2
}\,(4\,n\,e^{2}\,\sqrt{{\displaystyle \frac {\mathrm{f_n}}{(n - 2
)\,l^{2}}} } + l^{2} + l^{2}\,\sqrt{{\displaystyle \frac {
\mathrm{f_n}}{(n - 2)\,l^{2}}} })\,V\,\sqrt{{\displaystyle
\frac {n - 2}{n}} }  \nonumber \\ &&
(1 + \sqrt{{\displaystyle \frac {\mathrm{f_n}}{(n - 2)\,l^{2}}} }
)^{(n - 3)}\,(n - 1)\,n\,e \left/ {\vrule
height0.84em width0em depth0.84em} \right. \!  \! (l^{2}\,
\mathrm{f_n}\,\sqrt{{\displaystyle \frac {\mathrm{f_n}}{(n - 2)\,
l^{2}}} }), \nonumber \\
{\mathit{g\ }_{\mathit{ee}}}&=& - {\displaystyle \frac {1}{2}}
\sqrt{2}\,(4\,n\,e^{2}\,\sqrt{{\displaystyle \frac {\mathrm{f_n}
}{(n - 2)\,l^{2}}} } + l^{2} + l^{2}\,\sqrt{{\displaystyle
\frac {\mathrm{f_n}}{(n - 2)\,l^{2}}} })\,V\,\sqrt{
{\displaystyle \frac {n - 2}{n}} }  \nonumber \\ &&
(1 + \sqrt{{\displaystyle \frac {\mathrm{f_n}}{(n - 2)\,l^{2}}} }
)^{(n - 3)}\,(n - 1)\,n \left/ {\vrule
height0.84em width0em depth0.84em} \right. \!  \! (\sqrt{
{\displaystyle \frac {\mathrm{f_n}}{(n - 2)\,l^{2}}} }\,\mathrm{f_n}\,l).
\end{eqnarray} \end{scriptsize}

\begin{figure}
\hspace*{0.5cm}
\includegraphics[width=15.0cm,angle=-0]{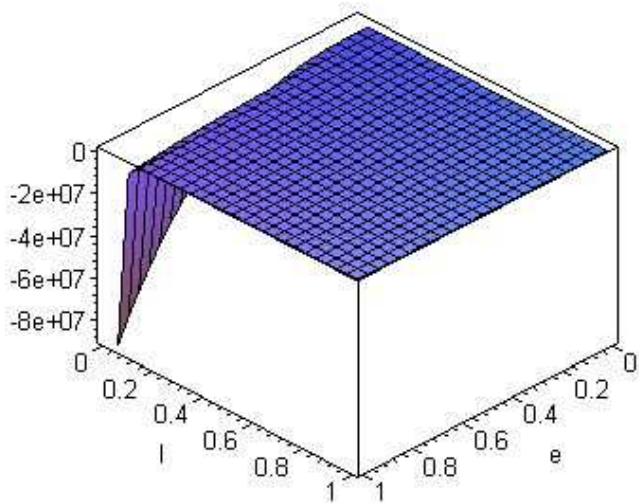}  \vspace*{-5.5cm}
\caption{The $ee$ component of Ruppeiner metric tensor plotted as
the function of $\{ e, l \}$, describing the fluctuations in six
dimensional topological Einstein-Yang-Mills black hole
configurations.} \label{ndree} \vspace*{-0.5cm}
\end{figure}

\begin{figure}
\hspace*{0.5cm}
\includegraphics[width=15.0cm,angle=-0]{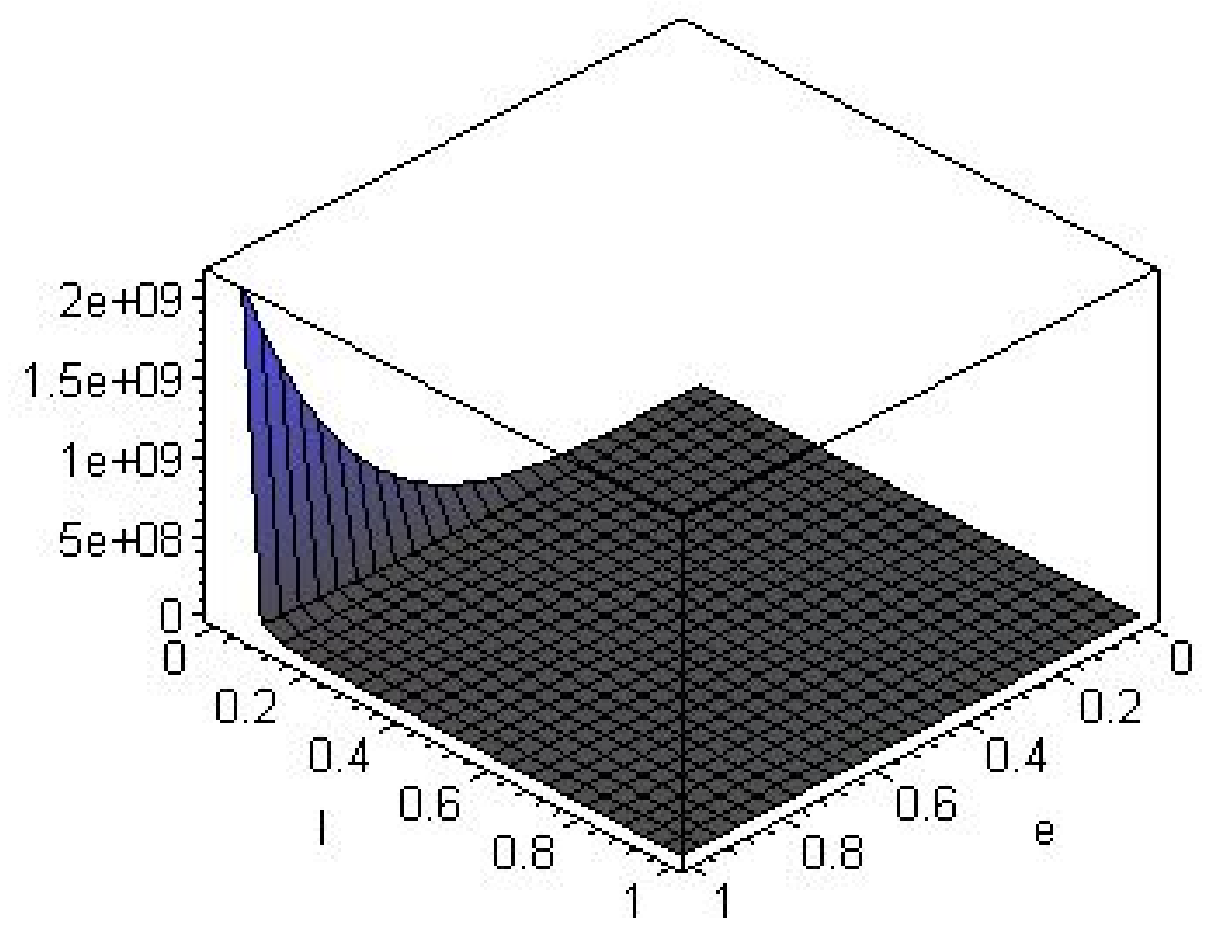}  \vspace*{-5.5cm}
\caption{The $el$ component of Ruppeiner metric tensor plotted as
the function of $\{ e, l \}$, describing the fluctuations in six
dimensional topological Einstein-Yang-Mills black hole
configurations.} \label{ndrel} \vspace*{-0.5cm}
\end{figure}

\begin{figure}
\hspace*{0.5cm}
\includegraphics[width=15.0cm,angle=-0]{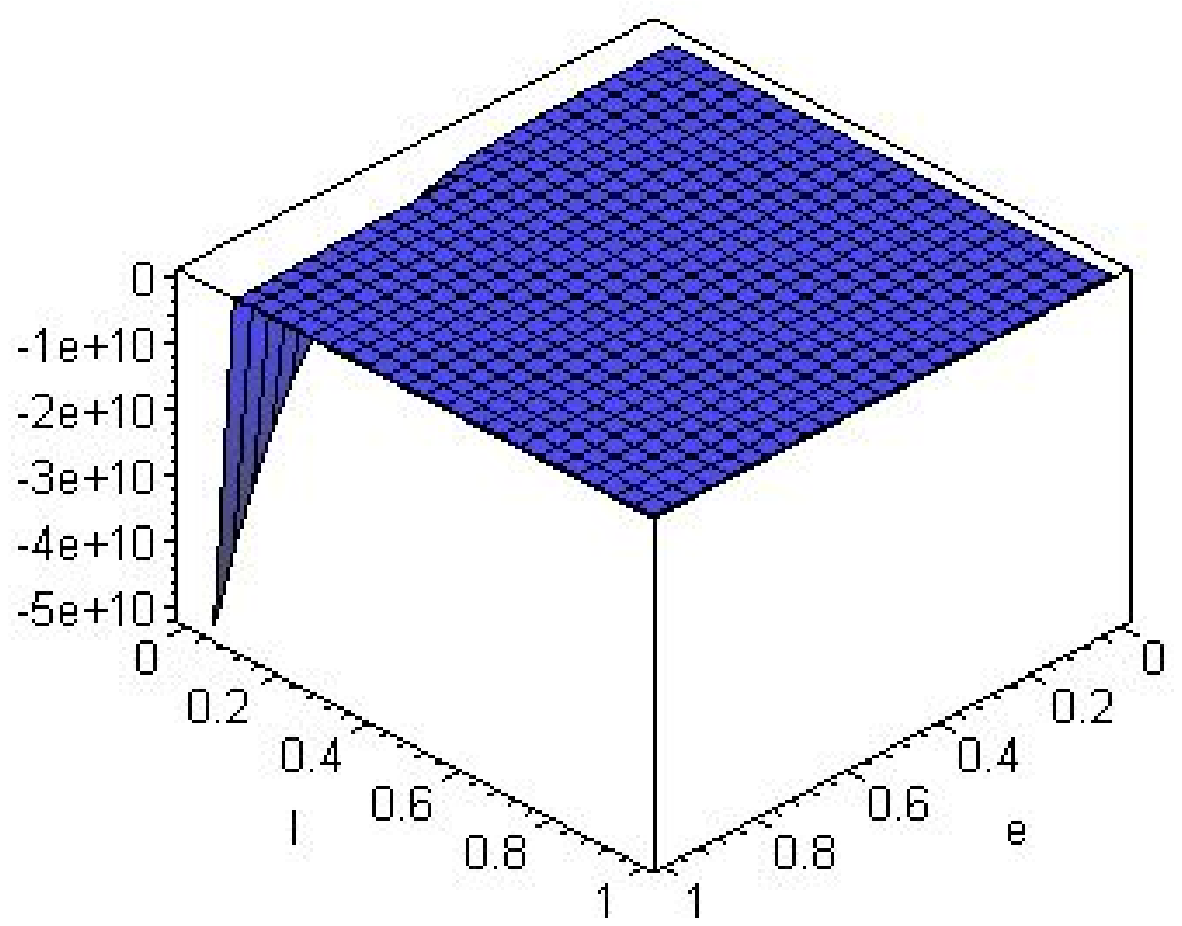}  \vspace*{-5.5cm}
\caption{The $ll$ component of Ruppeiner metric tensor plotted as
the function of $\{ e, l \}$, describing the fluctuations in six
dimensional topological Einstein-Yang-Mills black hole
configurations.} \label{ndrll} \vspace*{-0.5cm}
\end{figure}

Notice that the local stability characteristic of the higher dimensional
topological Einstein-Yang-Mills black holes follows from the positivity
of the heat capacities $\{g_{ee}, g_{ll} \}$ of the state-space metric tensor.
These are basically the diagonal components of the metric tensor associated
with entropy maximization of a chosen ensembl of the higher dimensional
topological Einstein-Yang-Mills black holes. For the choice of $V= 1$,
the explicit graphical view of the above mentioned local fluctuations
is depicted in the Figs.(\ref{ndree}, \ref{ndrll}). In the regime of
$e, l \in (0, 1)$, we see that the amplitude of $\{g_{ee} \}$ takes a value of
the order $-8 \times 10^{+07}$. In this range of the parameters $\{e, l\}$,
we find that the component $\{ g_{ll} \}$ lies in the range of $(-5 \times 10^{+10}, 0)$.
In this case, we observe that the growth of the amplitude of $\{g_{ee}, g_{ll} \}$
happens in the same distinct limit of $\{e, l\}$, that is a large $e$ and a small $l$.
In the above regime, for a given value of $e$ upto $0.15$, the system is nearly
stable upto in the range of $l \in (0.2, 1)$. Smaller values of the $l$ tend the
system towards an instability. The entropy of a large $e$ and a small $l$ leads
to an intrinsic state-space instability and thus can be the cause the black hole to
disappear. This is also forbidden by the blacl hole remanant hypothesis.
As shown in the Eqn. (\ref{metricrupn}), the entropy flow, namely the heat
capacities depends on the black hole parameters, and thus changing the value of a
parameter or the fluctuation in $\{e,l\}$ can affect the stability charecteristics
of the chosen higher dimensional topological Einstein-Yang-Mills black hole ensemble.
Thus, the diagonal component of the state-space metric tensor should be positive
for which the fluctuations provide a set of values of $\{e, l\}$ from which one can
determine the required local values of the flow parameters $e$ and $l$.
This signifies that the entropy extrimization could chareterize the underlying
thermodynamic instability of an ensemble of chosen black holes. Here, one is only
required to chose a specific domain of the parameters $\{e, l\}$ such that the
desired system remain in a well balanced limit of the entropy flow parameters.
Further, we notice from the Fig.(\ref{ndrel}) that the mix component $\{ g_{el}\}$
of the state-space metric tensor has a positive value under the entropy fluctuation.
Interestingly, we find that all the local fluctuations happen in a
small limit of the flow parameter $l$, and a large limit of the flow parameter $e$,
where higher value of $e$ happen to cause an instability. The Fig.(\ref{ndrel}) shows
that the higher dimensional topological Einstein-Yang-Mills black holes cannot flow
throught the vacuum, as long as there is a positive local heat capacity. In this examination,
the other parameters have to be kept constant in order to determine the limiting parametric
stability of the higher dimensional topological Einstein-Yang-Mills black hole ensemble.
In this sense, we see that the Figs.(\ref{ndree}, \ref{ndrll}, \ref{ndrel})
illustrate the local fluctuation properties of ensemble of arbitrary dimensional
topological Einstein-Yang-Mills black holes under the entropic flow of the $\{e, l\}$.
In fact, both the self pair fluctuations involving $\{e, l\}$, as defined by the metric tensor
$\{g_{ij}\ |\ i, j= e, l \}$ have only the negative numerical values, while the
the mix component $\{ g_{el}\}$ does not. More precisely, in order to see the global stability limit,
we require that the determinant of the metric tensor should be also positive in a chosen domain of
the parameters $\{e, l\}$. Thus, the values from a given set of fluctuating $\{e, l\}$
as illustrated in the Figs.(\ref{ndree}, \ref{ndrll}, \ref{ndrel}), the illustration
of the above type of ensemble of higher dimensional topological Einstein-Yang-Mills black holes
happens to be true locally. This is because of the fact that the determinant of the
state-space metric tensor vanishes identically for all values of  $\{e, l\}$.
Thus, the entropy extremization of higher dimensional topological Einstein-Yang-Mills black hole
ensemble for given $\{e, l\}$, in order to find a positive determinant regime,
would require further higher derivative stringy corrections or quantum loop corrections
to the entropy in order to keep an ensemble of topological Einstein-Yang-Mills black holes
globally stable.

In this case, since the determinant of the state-space metric tensor vanishes
identically for all values of the parameters $\{e, l\}$, and thus there is no question of
computing the state-space scalar curvature. From the perspective of intrinsic geometry,
we find that the Christofell symbols of arbitray topological Einstein-Yang-Mills
black hole can be expressed as the following expressions

\begin{scriptsize} \begin{eqnarray}
{\Gamma _{\mathit{eee}}}&=& -\frac{(n - 1)\,V
\sqrt{2}\,\sqrt{{\displaystyle \frac {n - 2}{n}} }\,(1 + \sqrt{
{\displaystyle \frac {\mathrm{f_n}}{(n - 2)\,l^{2}}} })^{(n - 4)}
\,n^{2}\,e}{(\sqrt{{\displaystyle \frac {\mathrm{f_n}}{(n - 2
)\,l^{2}}} }\,l^{3}\,\mathrm{f_n}^{2}\,(n - 2))}
( {\Gamma _{\mathit{eee}}}^{(1)}+
\,\sqrt{{\displaystyle \frac {\mathrm{f_n}}{(n - 2)\,l^{2}}} }
{\Gamma _{\mathit{eee}}}^{(2)}), \nonumber \\
{\Gamma _{\mathit{eel}}}&=& {\displaystyle \frac {1}{\sqrt{2}}}
\frac{V\,\sqrt{{\displaystyle \frac {n - 2
}{n}} }\,(1 + \sqrt{{\displaystyle \frac {\mathrm{f_n}}{(n - 2)\,
l^{2}}} })^{(n - 4)}\,(n - 1)\,n }
{(\sqrt{{\displaystyle \frac {\mathrm{f_n}}{(n - 2)\,l^{2}}} }\,l^{4}
\mathrm{f_n}^{2}\,(n - 2))}
({\Gamma _{\mathit{eel}}}^{(1)}+  \,\sqrt{{\displaystyle \frac {\mathrm{f_n}}{(n
 - 2)\,l^{2}}} }\, {\Gamma _{\mathit{eel}}}^{(2)}), \nonumber \\
{\Gamma _{\mathit{ele}}}&=& {\displaystyle \frac {1}{2}}
\frac{V\,\sqrt{2}\,\sqrt{{\displaystyle \frac {n - 2
}{n}} }\,(1 + \sqrt{{\displaystyle \frac {\mathrm{f_n}}{(n - 2)\,
l^{2}}} })^{(n - 4)}\,(n - 1)\,n}
{(\sqrt{{\displaystyle \frac {\mathrm{f_n}}{(n - 2)\,l^{2}}} }\,l^{4}
\mathrm{f_n}^{2}\,(n - 2))}
({\Gamma _{\mathit{ele}}}^{(1)}+\,\sqrt{{\displaystyle \frac {\mathrm{f_n}}{(n - 2)\,l^{2}}} }\,
{\Gamma _{\mathit{ele}}}^{(2)}), \nonumber \\
{\Gamma _{\mathit{ell}}}&=& - \frac{(n - 1)\,V\,\sqrt{2}\,\sqrt{{\displaystyle
\frac {n - 2}{n}} }\,(1 + \sqrt{{\displaystyle \frac {\mathrm{f_n
}}{(n - 2)\,l^{2}}} })^{(n - 4)}\,n\,e }
{(\sqrt{{\displaystyle \frac {\mathrm{f_n}}{(n - 2)\,l^{2}}} }\,l^{5}
\mathrm{f_n}^{2}\,(n - 2))}({\Gamma _{\mathit{ell}}}^{(1)}
+\,\sqrt{{\displaystyle \frac {\mathrm{f_n}}{(n - 2)\,l^{2}}} }\,
{\Gamma _{\mathit{ell}}}^{(2)}), \nonumber \\
{\Gamma _{\mathit{lle}}}&=& - \frac{(n - 1)\,V\,\sqrt{2}\,\sqrt{{\displaystyle
\frac{n - 2}{n}} }\,(1 + \sqrt{{\displaystyle \frac {\mathrm{f_n
}}{(n - 2)\,l^{2}}} })^{(n - 4)}\,n\,e}
{(\sqrt{{\displaystyle \frac {\mathrm{f_n}}{(n - 2)\,l^{2}}} }\,l^{5}  \mathrm{f_n}^{2}\,(n - 2)}
({\Gamma _{\mathit{lle}}}^{(1)}+ \,\sqrt{{\displaystyle \frac {\mathrm{f_n}}{(n - 2)\,l^{2}}} }\,
{\Gamma _{\mathit{lle}}}^{(2)}),  \nonumber \\
{\Gamma _{\mathit{lll}}}&=& {\displaystyle \frac {1}{2}}
\frac{V\,\sqrt{2}\,\sqrt{{\displaystyle
\frac {n - 2}{n}} }\,(1 + \sqrt{{\displaystyle \frac {\mathrm{f_n
}}{(n - 2)\,l^{2}}} })^{(n - 4)}\,(n - 1)\,n\,e^{2}}
{(\sqrt{{\displaystyle \frac {\mathrm{f_n}}{(n - 2)\,l^{2}}} }
\,l^{6}\,\mathrm{f_n}^{2}\,(n - 2))}
({\Gamma _{\mathit{lll}}}^{(1)}+ \,\sqrt{{\displaystyle \frac {\mathrm{f_n}}{(n
- 2)\,l^{2}}} } {\Gamma _{\mathit{lll}}}^{(2)}),
\end{eqnarray} \end{scriptsize}

where the factors $\mathit{\{ {\Gamma _{\mathit{ijk}}}^{(1)},
{\Gamma _{\mathit{ijk}}}^{(2)}| i,j,k \in \{e, l\} \}}$ of
arbitrary Ruppeiner metric tensor are given as

\begin{scriptsize} \begin{eqnarray}
{\Gamma _{\mathit{eee}}}^{(1)}&=&
16\,e^{4}\,n^{3}
- 48\,n^{2}\,e^{4}
+ 4\,e^{2}\,n^{3}\,l^{2}
- 8\,l^{2}\,n^{2}\,e^{2} \nonumber \\ &&
- 12\,l^{2}\,n\,e^{2}
+ 3\,l^{4}\,n^{2}
- 18\,l^{4}\,n
+ 24\,l^{4}, \nonumber \\
{\Gamma _{\mathit{eee}}}^{(2)}&=&
+ 3\,l^{4}\,n^{2}
- 18\,l^{4}\,n
+ 24\,l^{4}, \nonumber \\
{\Gamma _{\mathit{eel}}}^{(1)}&=&
32\,e^{6}\,n^{4}
- 64\,e^{6}\,n^{3}
+ 8\,e^{4}\,n^{4}\,l^{2}
- 48\,e^{4}\,l^{2}\,n^{2}   \nonumber \\ &&
+ 8\,e^{2}\,n^{3}\,l^{4}
- 38\,e^{2}\,l^{4}\,n^{2}
+ 44\,e^{2}\,l^{4}\,n
+ l^{6}\,n^{2}  \nonumber \\ &&
- 4\,n\,l^{6}
+ 4\,l^{6}, \nonumber \\
{\Gamma _{\mathit{eel}}}^{(2)}&=&
 8\,l^{2}\,n^{3}\,e^{4}
- 16\,e^{4}\,n^{2}\,l^{2}
+ 8\,l^{4}\,n^{3}\,e^{2}
- 40\,l^{4}\,n^{2}\,e^{2}  \nonumber \\ &&
+ 48\,l^{4}\,n\,e^{2}
+ l^{6}\,n^{2}
- 4\,l^{6}\,n
+ 4\,l^{6}, \nonumber \\
{\Gamma _{\mathit{ele}}}^{(1)}&=&
32\,e^{6} \,n^{4}
- 64\,e^{6}\,n^{3}
+ 8\,e^{4}\,n^{4}\,l^{2}
- 48\,e^{4}\,l^{2}\,n^{2}  \nonumber \\ &&
+ 8\,e^{2}\,n^{3}\,l^{4}
- 38\,e^{2}\,l^{4}\,n^{2}
+ 44\,e^{2}\,l^{4}\,n
+ l^{6}\,n^{2}  \nonumber \\ &&
- 4\,n\,l^{6}
+ 4\,l^{6},  \nonumber \\
{\Gamma _{\mathit{ele}}}^{(2)}&=&
8\,l^{2}\,n^{3}\,e^{4}
- 16\,e^{4}\,n^{2}\,l^{2}
+ 8\,l^{4}\,n^{3}\,e^{2}
- 40\,l^{4}\,n^{2}  \nonumber \\ &&
+ 48\,l^{4}\,n\,e^{2}
+ l^{6}\,n^{2}
- 4\,l^{6}\,n
+ 4\,l^{6}, \nonumber \\
{\Gamma _{\mathit{ell}}}^{(1)}&=&
16\,e^{6}\,n^{4}
- 16\,e^{6}\,n^{3}
+ 4\,e^{4}\,n^{4}\,l^{2}
+ 8\,e^{4}\,n^{3}\,l^{2}   \nonumber \\ &&
- 36\,e^{4}\,l^{2}\,n^{2}
+ 5\,e^{2}\,n^{3}\,l^{4}
- 20\,e^{2}\,l^{4}\,n^{2}
+ 20\,e^{2}\,l^{4}\,n  \nonumber \\ &&
+ l^{6}\,n^{2}
- 4\,n\,l^{6}
+ 4\,l^{6},  \nonumber \\
{\Gamma _{\mathit{ell}}}^{(2)}&=&
8\,l^{2}\,n^{3}\,e^{4}
- 16\,e^{4}\,n^{2}\,l^{2}
+ 5\,l^{4}\,n^{3}\,e^{2}
- 22\,l^{4}\,n^{2}\,e^{2} \nonumber \\ &&
+ 24\,l^{4}\,n\,e^{2}
+ l^{6}\,n^{2}
- 4\,l^{6}\,n
+ 4\,l^{6},  \nonumber \\
{\Gamma _{\mathit{lle}}}^{(1)}&=&
16\,e^{6}\,n^{4}
- 16\,e^{6}\,n^{3}
 + 4\,e^{4}\,n^{4}\,l^{2}
+ 8\,e^{4} \,n^{3}\,l^{2}  \nonumber \\ &&
- 36\,e^{4}\,l^{2}\,n^{2}
+ 20\,e^{2}\,l^{4}\,n
+ l^{6}\,n^{2}
- 4\,n\,l^{6}   \nonumber \\ &&
+ 4\,l^{6}
+ 5\,e^{2}\,n^{3}\,l^{4}
- 20\,e^{2}\,l^{4}\,n^{2},  \nonumber \\
{\Gamma _{\mathit{lle}}}^{(2)}&=&
8\,l^{2}\,n^{3}\,e^{4}
- 16\,e^{4}\,n^{2}\,l^{2}
+ 5\,l^{4}\,n^{3}\,e^{2}
- 22\,l^{4}\,n^{2}\,e^{2}\nonumber \\ &&
+ 24\,l^{4}\,n\,e^{2}
+ l^{6}\,n^{2}
- 4\,l^{6}\,n
 + 4\,l^{6}, \nonumber \\
{\Gamma _{\mathit{lll}}}^{(1)}&=&
32\,e^{6} \,n^{4}
+ 8\,e^{4}\,n^{4}\,l^{2}
+ 32 \,e^{4}\,n^{3}\,l^{2}
- 96\,e^{4}\,l^{2}\,n^{2}   \nonumber \\ &&
+ 12\,e^{2}\,n^{3}\,l^{4}
- 42\,e^{2}\,l^{4}\,n^{2}
+ 36\,e^{2}\,l^{4}\,n
+ 3\,l^{6}\,n^{2}  \nonumber \\ &&
- 12\,n\,l^{6}
 + 12\,l^{6}, \nonumber \\
{\Gamma _{\mathit{lll}}}^{(2)}&=&
24\,l^{2}\,n^{3}\,e^{4}
- 48\,e^{4}\,n^{2}\,l^{2}
+ 12\,l^{4}\,n^{3}\,e^{2}
- 48\,l^{4}\,n^{2}\,e^{2} \nonumber \\ &&
+ 48\,l^{4}\,n\,e^{2}
+ 3\,l^{6}\,n^{2}
- 12\,l^{6}\,n
+ 12\,l^{6}. \nonumber \\
\end{eqnarray} \end{scriptsize}

\section{Weinhold Geometry}
In this section, we illustrate the role of thermodynamic geometry
from the perspective of the energy minimization, that is here the
minimization of the ADM mass of the topological
Einstein-Yang-Mills black hole in spacetime dimension $D=5$. Given
an ensemble of such black hole configurations, the optimization of
the energy is necessary in order to define the vacuum stability of
the Yang-Mills gauge theory with a non-trivial topological black
hole, where not only the field theory is considered to have a
flauctuting vacuum but also an imminent black have in the
spacetime background of the Yang-Mills vacuum ensemble. To
illustrate the hypotheis, we shall first consider the case of five
dimensional theory and then in the next subsection generalize it
to an ensemble of arbitary dimensional topological
Einstein-Yang-Mills black holes. It is worth mentioning that the
energy  minimization is intrinsically same the entropy
maximization up to a Legendre transformation. Thus, we may notice
that the flow equations and the analysis of the parametric
stability constraint in the sense of the Weinhold geometry follows
directly up to a sign with the definition of the Ruppeiner
geometry. Both of these thermodynamic geometries are intertuned
each other, as we have defined then in the section $2$.

\subsection{Five Dimensional Black Holes}

From the Ref. \cite{BD}, the ADM mass of a topological Einstein-Yang-Mills
black hole in spacetime dimension $D=5$ is given by

\begin{scriptsize} \begin{eqnarray}
\mathrm{M}(e, \,l) :=  - {\displaystyle \frac {1}{3}} \,e^{2}\,(
\mathrm{ln}({\displaystyle \frac {1}{4}} \,l^{2} +
{\displaystyle \frac {1}{4}} \,l\,(l^{2} + 8\,e^{2})) -
{\displaystyle \frac {1}{2}} ) - {\displaystyle \frac {1}{24}} \,
l^{2} - {\displaystyle \frac {1}{24}} \,l\,(l^{2} + 8\,e^{2}).
\end{eqnarray} \end{scriptsize}

With the notions of the thermodynamic geometry, the Legendre
associated Weinhold line element can be expressed as

\begin{scriptsize} \begin{eqnarray}
\mathit{\ ds}^{2}=({\frac {\partial ^{2}}{\partial e^{2}}}\,
\mathrm{M}(e, \,l))\,\mathit{\ d}\,e^{\mathit{2\ }} + 2\,(
{\frac {\partial ^{2}}{\partial l\,\partial e}}\,\mathrm{M}(e, \,
l))\,\mathit{\ d}\,e^{\ }\,\mathit{d\ }\,l^{\ } + ({\frac {
\partial ^{2}}{\partial l^{2}}}\,\mathrm{M}(e, \,l))\,\mathit{\ d
}\,l^{\mathit{2\ }}.
\end{eqnarray} \end{scriptsize}

To simplify the expression of the determinant of the metric tensor,
the appropriate scaling function turns out to be

\begin{scriptsize} \begin{eqnarray}
\mathrm{b} := l + l^{2} + 8\,e^{2}.
\end{eqnarray} \end{scriptsize}

For the above given mass, the computation of the Hessian matrix
shows the following expressions for the components of the Weinhold
metric tensor

\begin{scriptsize} \begin{eqnarray}
{\mathit{g\ }_{\mathit{ee}}} &=& {\displaystyle \frac {1}{3\, \mathrm{b}^{2}}}
( -320 \,e^{4} - 64\,l\,e^{2} - 96\,l^{2}\,e^{2} - 32\,l^{3}\,e^{2} -
128\,l\,e^{4} - 3\,l^{4} \nonumber \\ && + 64\,\mathrm{ln}(2)\,l\,e^{2
} + 4\,\mathrm{ln}(2)\,l^{2} + 8\,\mathrm{ln}(2)\,l^{3} + 4 \,\mathrm{ln}(2)\,l^{4}
+ 256\,\mathrm{ln}(2)\,e^{4} \nonumber \\ && + 64\,\mathrm{ln}(2)\,l^{2}\,e^{2}
- 2\,\mathrm{ln}(l\,\mathrm{b})\,l^{4} - 128\,\mathrm{ln}(l\,\mathrm{b})\,e^{4}
- 32\,\mathrm{ln}(l\,\mathrm{b})\,l\,e^{2} \nonumber \\ && - 2\,l^{5}
- 2\,\mathrm{ln}(l\,\mathrm{b})\,l^{2} - 4\,\mathrm{ln}(l\,\mathrm{b})\,l^{3}
- 32\,\mathrm{ln}(l\,\mathrm{b})\,l^{2}\,e^{2} + l^{2}), \nonumber  \\
{\mathit{g\ }_{\mathit{el}}} &=& - {\displaystyle \frac {2\,e}{3\, \mathrm{b}^{2}\, l }} \,
{\displaystyle (2\,l^{2} + 6\,l^{3} + 16\,l\,e^{2} + 5 \,l^{4} + 32\,l^{2}\,e^{2}
+ 64\,e^{4}} \nonumber \\ && {\displaystyle + l^{5} + 16\,l^{3}\,e^{2}
+ 64\,l\,e^{4})}, \nonumber  \\
{\mathit{g\ }_{\mathit{ll}}}&=& {\displaystyle \frac {1}{12\, \mathrm{b}^{2}\, l^{2}}} \,
{\displaystyle (8\,l^{2}\,e^{2} + 64\,l\,e^{4} - 52\,l^{4}
\,e^{2} - 64\,e^{4}\,l^{2} + 256\,e^{6}} \nonumber \\ && {\displaystyle
- l^{4} - 5\,l^{5} - 7\,l^{6} - 3\,l^{7} - 48\,l^{5}\,e^{2} - 192\,l^{3}\,e^{4})}.
\end{eqnarray} \end{scriptsize}

\begin{figure}
\hspace*{0.5cm}
\includegraphics[width=15.0cm,angle=-0]{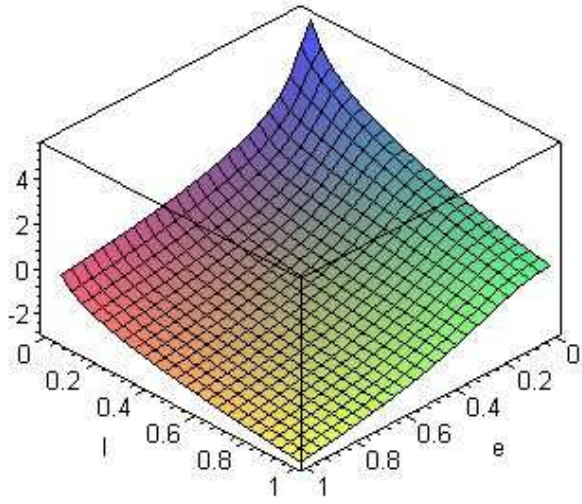}  \vspace*{-5.5cm}
\caption{The $ee$ component of Weinhold metric tensor plotted as
the function of $\{ e, l \}$, describing the fluctuations in five
dimensional topological Einstein-Yang-Mills black hole
configurations.} \label{5dwee} \vspace*{-0.5cm}
\end{figure}

\begin{figure}
\hspace*{0.5cm}
\includegraphics[width=15.0cm,angle=-0]{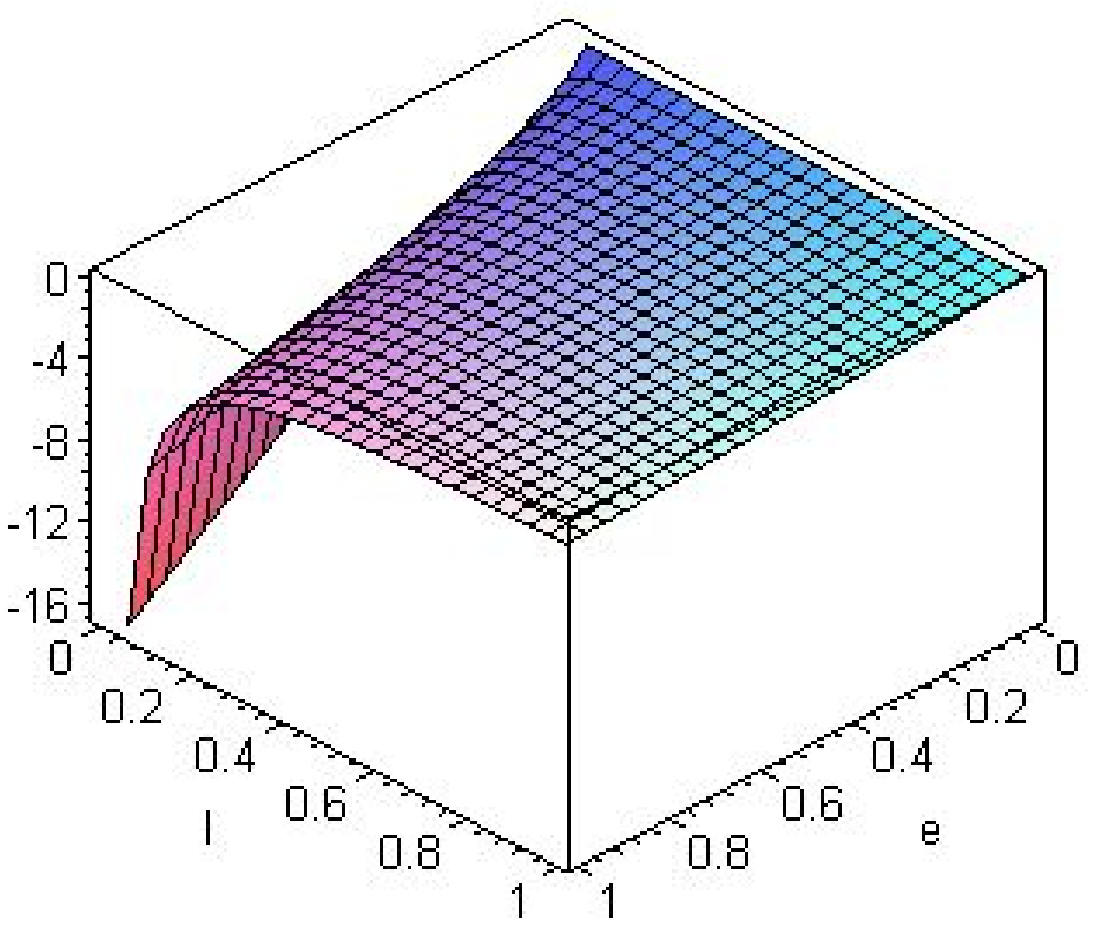}  \vspace*{-5.5cm}
\caption{The $el$ component of Weinhold metric tensor plotted as
the function of $\{ e, l \}$, describing the fluctuations in five
dimensional topological Einstein-Yang-Mills black hole
configurations.} \label{5dwel} \vspace*{-0.5cm}
\end{figure}

\begin{figure}
\hspace*{0.5cm}
\includegraphics[width=15.0cm,angle=-0]{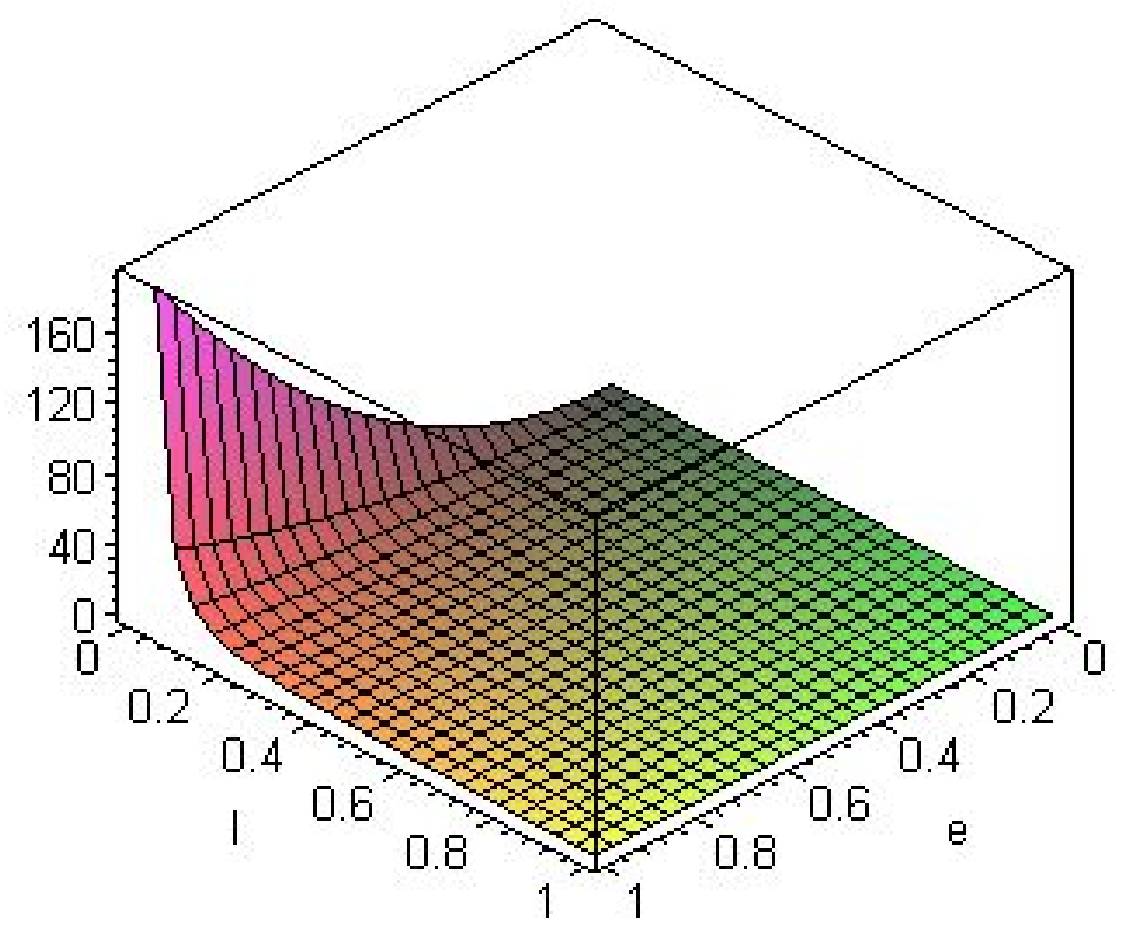}  \vspace*{-5.5cm}
\caption{The $ll$ component of Weinhold metric tensor plotted as
the function of $\{ e, l \}$, describing the fluctuations in five
dimensional topological Einstein-Yang-Mills black hole
configurations.} \label{5dwll} \vspace*{-0.5cm}
\end{figure}

In this case, we see that the heat capacities have rather diverse
charectors. For the case of $V= 1$, the heat capacities $\{g_{ee},
g_{ll} \}$ are shown in the Figs.(\ref{5dwee}, \ref{5dwll}).
Herewith, in the interval of $e, l \in (0, 1)$, the amplitude of
$\{g_{ll} \}$ takes a positive value of the order $200$. In this
range of the parameters $\{e, l\}$, Fig.(\ref{5dwee}) shows that
the component $\{ g_{ee} \}$ lies in the range of $(-4, +6)$. In
this case, we hereby observe that the fluctuations of both the
$\{g_{ee}, g_{ll} \}$ do not occure with a positive amplitude of
the fluctuations. Namely, the $ee$ component fluctuations are
generically present near the origin of the flow parameters, while
this is not the case for the $ll$ component. We see that the  $ll$
component energy fluctuations are largly present for a large $e$
and a small $l$. The Fig.(\ref{5dwel}) shows that the
corresponding mix component of the complex ADM mass flow
fluctuation, in which the $el$component of the Weinhold metric
takes a maximum amplitude of the order $-16$. The above plots may
change for a different vacuum black hole and for a different field
theory, as well. The values of $e$ and $l$ are sensitive to higher
derivative and higher order corrections as well. This analysis can
further be extend for a different Lagrangian of the theory and the
background spacetime black hole ensemble.

\begin{figure}
\hspace*{0.5cm}
\includegraphics[width=15.0cm,angle=-0]{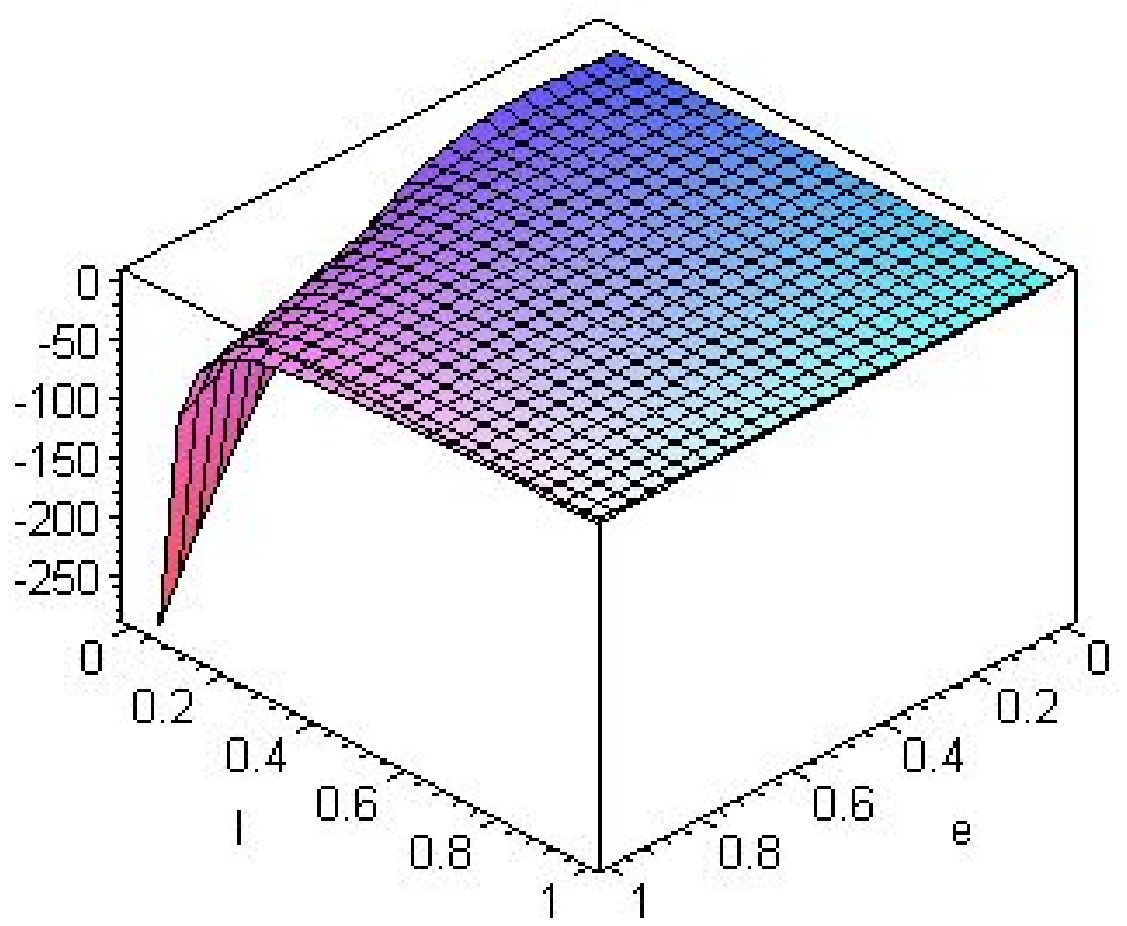}  \vspace*{-5.5cm}
\caption{The determinant of Weinhold metric tensor plotted as the
function of $\{ e, l \}$, describing the fluctuations in five
dimensional topological Einstein-Yang-Mills black hole
configurations.} \label{5dwdetg} \vspace*{-0.5cm}
\end{figure}

\begin{figure}
\hspace*{0.5cm}
\includegraphics[width=15.0cm,angle=-0]{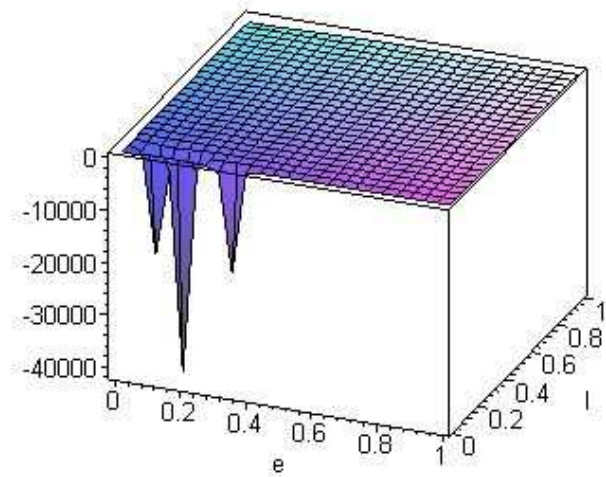}  \vspace*{-5.5cm}
\caption{The Weinhold curvature scalar plotted as the function of
$\{ e, l \}$, describing the fluctuations in five dimensional
topological Einstein-Yang-Mills black hole configurations.}
\label{5dwR} \vspace*{-0.5cm}
\end{figure}

To illustate the metric structure properties of the above mass
fluctuations, we now offer the fluctuation properties of the
metric tensor $\mathit{g_{ij}}$. For a given intrinsic Weinhold
surface of $\mathit{\{e, l\}}$, these fluctuations are precisely
depicted by the following Christofel symbols

\begin{scriptsize} \begin{eqnarray}
{\Gamma _{\mathit{eee}}} &=& - {\displaystyle \frac {32}{3\, \mathrm{b}^{3}}} \,
{\displaystyle (e\,(3\,l^{2} + 6\,l^{3} + 12\,l\,e^{2} + 3
\,l^{4} + 12\,l^{2}\,e^{2} + 32\,e^{4}))}, \nonumber  \\
{\Gamma _{\mathit{eel}}} &=& - {\displaystyle \frac {1}{3\,l\,\mathrm{b}^{3}}}
(2\,l^{3} + 8\,l^{4} + 11\,l^{5} + 192\,l\,e^{4}
+ 512\,l\,e^{6} + 24\,l^{5}\,e^{2} \nonumber \\ &&
+ 192\,l^{3}\,e^{4} + l^{7} + 6\,l^{6} + 512\,e^{6}
+ 24\,l^{4}\,e^{2} + 384\,e^{4}\,l^{2}), \nonumber  \\
{\Gamma _{\mathit{ele}}} &=& - {\displaystyle \frac {1}{3\,l\,\mathrm{b}^{3}}}
(2\,l^{3} + 8\,l^{4} + 11\,l^{5} + 192\,l\,e^{4} + 512\,l\,e^{6} + 24\,l^{5}\,e^{2} \nonumber \\ &&
+ 192\,l^{3}\,e^{4} + l^{7} + 6\,l^{6} + 512\,e^{6}+ 24\,l^{4}\,e^{2} + 384\,e^{4}\,l^{2}), \nonumber  \\
{\Gamma _{\mathit{ell}}} &=& {\displaystyle \frac {1}{3\,l^{2}\,\mathrm{b}^{3}}} \,
{\displaystyle (e\,(2\,l^{3} + 6\,l^{4} + 16\,l^{2}\,e^{2}
+ 7\,l^{5} + 192\,l\,e^{4} + 3\,l^{6}} \nonumber \\ && {\displaystyle + 512\,e^{6} - 24\,l^{4}\,
e^{2} + 192\,e^{4}\,l^{2}))}, \nonumber  \\
{\Gamma _{\mathit{lle}}} &=& {\displaystyle \frac {1}{3\,l^{2}\,\mathrm{b}^{3}}} \,
{\displaystyle (e\,(2\,l^{3} + 6\,l^{4} + 16\,l^{2}\,e^{2}
+ 7\,l^{5} + 192\,l\,e^{4} + 3\,l^{6}} \nonumber \\ && {\displaystyle
+ 512\,e^{6} - 24\,l^{4}\, e^{2} + 192\,e^{4}\,l^{2}))}, \nonumber  \\
{\Gamma _{\mathit{lll}}} &=& - {\displaystyle \frac {1}{24\,l^{3}\,\mathrm{b}^{3}}}
(1536\,l\,e^{6} + 120\,l^{5}\,e^{2} + 192\,l^{3}\,e^{4} + 9\,l^{7} + 3\,
l^{6} \nonumber \\ && + 48\,l^{4}\,e^{2} + 16\,l^{3}\,e^{2} + 192\,e^{4}\,l^{2}
+ 168\,l^{6}\,e^{2} + 384\,l^{4}\,e^{4} \nonumber \\ && + 9\,l^{8} + 3\,l^{9}
+ 72\,l^{7}\,e^{2} + 576\,l^{5}\,e^{4} + 1536\,l^{3}\,e^{6}
+ 4096\,e^{8} \nonumber \\ && + 1536\,l^{2}\,e^{6}).
\end{eqnarray} \end{scriptsize}

Herewith, the determinant of the metric tensor reduces to the following formula

\begin{scriptsize} \begin{eqnarray}
\mathit{g\ }&=& {\displaystyle \frac {1}{36\, l^{2}\,\mathrm{b}^{3}}} \sum_{i=0}^4 r_i(l) e^{2\,i},
\end{eqnarray} \end{scriptsize}

where the coefficients $\mathit{r_i(l)}$ can be written as the following expressions

\begin{scriptsize} \begin{eqnarray}
r_0&=&
(2\,\mathrm{ln}(l\,\mathrm{b}) - 4\,\mathrm{ln}(2) -1) l^{5}
+4\,( 3\,\mathrm{ln}(l\,\mathrm{b})- 6\,\mathrm{ln}(2) - 1)\,l^{6} \nonumber \\ &&
+24\,(\mathrm{ln}(l\,\mathrm{b}) - 2\, \mathrm{ln}(2))\,l^{7}
+2\,(10\,\mathrm{ln}(l\,\mathrm{b})- 20\,\mathrm{ln}(2)+ 7)\,l^{8}  \nonumber \\ &&
+ (6\,\mathrm{ln}(l\,\mathrm{b}) - 12\,\mathrm{ln}(2) + 17)\,l^{9}
+ 6\,l^{10},  \nonumber \\
r_1&=&
-8\,(2\,\mathrm{ln}(l\,\mathrm{b})- 4\,\mathrm{ln}(2)+ 7)\,l^{3}
- 256\,l^{4}
+4\,(46\,\mathrm{ln}(l\,\mathrm{b}) - 92\,\mathrm{ln}(2) -67)\,l^{5}  \nonumber \\ &&
+4\,( 78\,\mathrm{ln}(l\,\mathrm{b}) - 156\,\mathrm{ln}(2)+  35)\,l^{6}
+16\,( 9\,\mathrm{ln}(l\,\mathrm{b}) - 18\,l^{7}\,\mathrm{ln}(2)+ 21)\,l^{7}  \nonumber \\ &&
+ 128\,l^{8},  \nonumber \\
r_2&=&
- 256\,(\mathrm{ln}(l\,\mathrm{b}) - 2\,\mathrm{ln}(2) + 4)\,l^{2}
- 2560\,l^{3} \nonumber \\ &&
+32\,( 42\,\mathrm{ln}(l\,\mathrm{b}) - 84\,\mathrm{ln}(2)- 19)\,l^{4}
+128\,( 9\,\mathrm{ln}(l\,\mathrm{b}) - 18\,\mathrm{ln}(2) + 14)\,l^{5} \nonumber \\ &&
+ 768\,l^{6}, \nonumber \\
r_3&=&
-256\,(6\,l\,\mathrm{ln}(l\,\mathrm{b}) - 12\,l\,\mathrm{ln}(2) + 25)\,l
+ 256\,(2\,\mathrm{ln}(l\,\mathrm{b}) - 4\,l^{2}\,\mathrm{ln}(2) - 35)\,l^{2} \nonumber \\ &&
+1024\,( 3\,\mathrm{ln}(l\,\mathrm{b})- 6\,l^{3}\,\mathrm{ln}(2) - 1)\,l^{3}, \nonumber \\
r_4&=& - 18432 + 8192\,\mathrm{ln}(2) - 20480\,l
- 4096\,\mathrm{ln}(l\,\mathrm{b}) - 8192\,l^{2}.
\end{eqnarray} \end{scriptsize}

As mentioned in the previous section for the fluctuating entropy
of an ensemble of five dimensional topological Einstein-Yang-Mills
black hole, we see in this case that the ensemble stability of the
fluctuating configuration can be determined in terms of the values
of the determinant of the Weinhold metric tensor, as the function
of the mass flow parameter $\{e, l\}$, as defined above.
Furthermore, the gloal behavior of the system  follows from the
determinant of fluctuation metric tensor. For the cases of $V=\pm
1$, we observe that the determinant of the Weinhold metric tensor
tends to a negative value. Namely, we see from the
Fig.(\ref{5dwdetg}) that the peak of the determinant of the
Weinhold metric tensor is of the order $-300$. As in the case of
the local energy fluctuations of the five dimensional topological
Einstein-Yang-Mills black holes, we find that the global energy
fluctuations also happen for a small value of the parameter $l$
and a large value of the charge $e$. When only one of the
parameter is allowed to vary, the stability of the the five
dimensional topological Einstein-Yang-Mills black hole enseble is
determined by the positivity of the first principle minor $p_1:=
g_{ee}$. Physically, the above qualitative demonstrations of the
energy fluctuations illustrate the parametric stability properties
of an ensemble of two parameter five dimensional topological
Einstein-Yang-Mills black holes.

With convention $\mathit{ \mathrm{\tilde{b}} := \mathrm{ln}(l\,(l + l^{2} + 8\,e^{2}))}$,
we find that the scalar curvature possesses the following non-trivial expression

\begin{scriptsize} \begin{eqnarray}
\mathit{R\ } &=& 12 \mathrm{\frac{\sum_{i=0}^7 R_i(l) e^{2i}}{(\sum_{i=0}^4 r_i(l) e^{2\,i})^{2}}},
\end{eqnarray} \end{scriptsize}

where the factors in the numerator are given by the expressions
\begin{scriptsize} \begin{eqnarray}
R_0&=&
- 8\,l^{8}
+ (12\,\mathrm{\tilde{b}} - 24\,\mathrm{ln}(2) - 94)\,l^{9}
+ (84\,\mathrm{\tilde{b}} - 168\,\mathrm{ln}(2) - 438)\,l^{10} \nonumber \\ &&
+ (246\,\mathrm{\tilde{b}} - 492\,\mathrm{ln}(2) - 1087)\,l^{11}
+ (390\,\mathrm{\tilde{b}} - 780\,\mathrm{ln}(2) - 1591)\,l^{12} \nonumber \\ &&
+ (360\,\mathrm{\tilde{b}} - 720\,\mathrm{ln}(2) - 1412)\,l^{13}
+ (192\,\mathrm{\tilde{b}} - 384\,\mathrm{ln}(2) - 740)\,l^{14}  \nonumber \\ &&
+ (54\,\mathrm{\tilde{b}} - 108\,\mathrm{ln}(2) - 207)\,l^{15}
+ (6\,\mathrm{\tilde{b}} - 12\,\mathrm{ln}(2) - 23)\,l^{16}, \nonumber \\
R_1&=&
- 64\,l^{6}
+ (64 \,\mathrm{\tilde{b}} -128\, \mathrm{ln}(2) - 928)\,l^{7}
+ (576 \,\mathrm{\tilde{b}} - 1152 \,\mathrm{ln}(2) - 5728)\,l^{8}  \nonumber \\ &&
+ (2304 \,\mathrm{\tilde{b}} - 4608 \,\mathrm{ln}(2)- 17984)\,l^{9}
+ (5008 \,\mathrm{\tilde{b}} - 10016 \,\mathrm{ln}(2) - 31976)\,l^{10}  \nonumber \\ &&
+ (6288 \,\mathrm{\tilde{b}} - 12576 \,\mathrm{ln}(2) - 33560)\,l^{11}
+ (4560 \,\mathrm{\tilde{b}} -9120 \,\mathrm{ln}(2) - 20448)\,l^{12} \nonumber \\ &&
+ (1776 \,\mathrm{\tilde{b}} - 3552 \,\mathrm{ln}(2) - 6584)\,l^{13}
+ (288 \,\mathrm{\tilde{b}} - 576 \,\mathrm{ln}(2)- 840)\,l^{14}, \nonumber \\
R_2&=&
(512 \,\mathrm{ln}(2) - 256 \,\mathrm{\tilde{b}} - 3456)\,l^{5}
+ (768 \,\mathrm{\tilde{b}} - 1536 \,\mathrm{ln}(2) - 27776)\,l^{6}   \nonumber \\ &&
+ (8704 \,\mathrm{\tilde{b}} - 17408 \,\mathrm{ln}(2) - 107776)\,l^{7}
+ (26880 \,\mathrm{\tilde{b}} - 53760 \,\mathrm{ln}(2) - 237952)\,l^{8}   \nonumber \\ &&
+ (45312 \,\mathrm{\tilde{b}} - 90624\,\mathrm{ln}(2)- 306432)\,l^{9}
+ (45312 \,\mathrm{\tilde{b}} - 90624 \,\mathrm{ln}(2) - 224128)\,l^{10}  \nonumber \\ &&
+ (24960 \,\mathrm{\tilde{b}} - 49920 \,\mathrm{ln}(2) - 84672)\,l^{11}
+ (5760 \,\mathrm{\tilde{b}} - 11520 \,\mathrm{ln}(2) - 12480)\,l^{12}, \nonumber \\
R_3&=&
+ (8192 \,\mathrm{ln}(2)- 4096 \,\mathrm{\tilde{b}} - 55296)\,l^{4}
+ (12288 \,\mathrm{\tilde{b}}- 24576 \,\mathrm{ln}(2) - 321536)\,l^{5}  \nonumber \\ &&
+ (98304 \,\mathrm{\tilde{b}} - 196608 \,\mathrm{ln}(2) - 892928)\,l^{6}
+ (221184 \,\mathrm{\tilde{b}} - 442368 \,\mathrm{ln}(2)- 1427456)\,l^{7}  \nonumber \\ &&
+ (276480 \,\mathrm{\tilde{b}} - 552960 \,\mathrm{ln}(2)- 1281536)\,l^{8}
+ (199680 \,\mathrm{\tilde{b}} - 399360 \,\mathrm{ln}(2) - 574976)\,l^{9} \nonumber \\ &&
+ (61440\,\mathrm{\tilde{b}} - 122880\,\mathrm{ln}(2) - 96768)\,l^{10}, \nonumber \\
R_4&=&
(49152 \,\mathrm{ln}(2) - 24576 \,\mathrm{\tilde{b}} - 430080)\,l^{3}
+ (155648\ \,\mathrm{\tilde{b}} - 311296\,\mathrm{ln}(2)- 1945600)\,l^{4} \nonumber \\ &&
+ (737280\,\mathrm{\tilde{b}} - 1474560\,\mathrm{ln}(2)  - 3956736)\,l^{5}
+ (1179648\,\mathrm{\tilde{b}}- 2359296\,\mathrm{ln}(2) - 4325376)\,l^{6} \nonumber \\ &&
+ (991232\,\mathrm{\tilde{b}} - 1982464\,\mathrm{ln}(2) - 2281472)\,l^{7}
+ (368640\,\mathrm{\tilde{b}} - 737280\,\mathrm{ln}(2) - 413696)\,l^{8}, \nonumber \\
R_5&=&
+ (131072 \,\mathrm{ln}(2) - 65536 \,\mathrm{\tilde{b}} - 1933312)\,l^{2}
+ (983040 \,\mathrm{\tilde{b}}- 1966080\,\mathrm{ln}(2)- 6586368)\,l^{3}  \nonumber \\ &&
+ (2949120 \,\mathrm{\tilde{b}}- 5898240 \,\mathrm{ln}(2) - 8978432)\,l^{4}
+ (2949120 \,\mathrm{\tilde{b}} - 5898240 \,\mathrm{ln}(2)- 5406720)\,l^{5} \nonumber \\ &&
+ (1179648 \,\mathrm{\tilde{b}} - 2359296\,\mathrm{ln}(2) - 884736)\,l^{6}, \nonumber \\
R_6&=&
- 4718592\,l
+ (3145728\,\mathrm{\tilde{b}} - 629145\,\mathrm{ln}(2) - 9961472)\,l^{2} \nonumber \\ &&
+ (4718592\,\mathrm{\tilde{b}}- 9437184\,\mathrm{ln}(2)- 6553600)\,l^{3}
+ (1572864\,\mathrm{\tilde{b}} - 3145728\,\mathrm{ln}(2) - 262144)\,l^{4}, \nonumber \\
R_7&=&
- 6291456
+ (4194304 \,\mathrm{\tilde{b}} - 8388608\,\mathrm{ln}(2)- 2097152)\,l
+ 2097152\,l^{2}.
\end{eqnarray} \end{scriptsize}

It is worth mentioning that the coefficients $\mathit{\{r_i(l)|\ i=0,1,2,3,4 \}}$,
appearing in the denominator of the scalar curvature, remain the functions
as those in the numerator of the determinant of the metric tensor.

In general, it is worth mentioning that the long rang correlation
are characterized as per the definition of the scalar curvature.
Namely, the global stability properties of the five dimensional
topological Einstein-Yang-Mills black hole enseble follow from the
corresponding thermodynamic scalar curvature. In particular,
for the range of $e,l \in (0, 1)$, the Fig.(\ref{5dwR}) shows that
the scalar curvature has a ngative amplitude of order $-40000$.
This shows that the underlying five dimensional topological
Einstein-Yang-Mills black hole configuration is an interacting system.
The negative sign of the scalar curvature signifies the attractive nature
of the statisical interactions. For the case of $e \in (0,1)$ and $l \in (0, 1)$,
we notice from the Fig.(\ref{5dwR}) that there exist a number of large negative peak
of the global interactions of the order $-20000$ to $-40000$. In this sense,
upto a small range of $e$ and $l$, the two parameter five dimensional
topological Einstein-Yang-Mills black holes behave as a stable configuration,
however an increasing value of $\{e,l\}$ cannot increase the limit of the parametric
stability, as it could make a negative value of the determinant of the metric tensor.
Thus, when the parameter $\{e, l\}$ are allowed to fluctuate, the above figure
Fig.(\ref{5dwR}) indicates that the interaction properties of the underlying
five dimensional topological Einstein-Yang-Mills black hole configuration.
The above instability analysis shows the above black hole system for small values
of the vacuum parameters is highly sensitivity to the statistical fluctuations.

\subsection{Higher Dimensional Black Holes}

From the Ref. \cite{BD}, the ADM mass of a topological Einstein-Yang-Mills
black hole in arbitrary spacetime dimension $D=1+n$ can be expressed by

\begin{scriptsize} \begin{eqnarray}
\mathrm{M}(e, \,l) &:=&  - \frac{2}{n-1} (\sqrt{{\displaystyle \frac {n - 2}{2\, n}} }\,
l\,\sqrt{1 + \sqrt{1 + {\displaystyle \frac {4\,n\,e^{2}}{(n - 2)\,l^{2}}} }})^{(n - 4)} \times
\nonumber \\ && \left(  \! {\displaystyle \frac {1}{4}} \,{\displaystyle \frac{(n - 2)\,l^{2}\,
(1 + \sqrt{1 + {\displaystyle \frac {4\,n\,e^{2}}{(n - 2)\,l^{2}}} })^{2}}{n^{2}}}
+ {\displaystyle \frac {e^{2}}{n - 4}}  \!  \right).
\end{eqnarray} \end{scriptsize}

In this case, let the scaling function be defined as

\begin{scriptsize} \begin{eqnarray}
\mathrm{b_n} := {\displaystyle \frac {l^{2}\,n - 2\,l^{2} + 4\,n
\,e^{2}}{(n - 2)\,l^{2}}}.
\end{eqnarray} \end{scriptsize}

Thence, for a given mass of the five dimensional extremal
topological Einstein-Yang-Mills black hole, we find the following
components of Weinhold metric tensor

\begin{scriptsize} \begin{eqnarray} \label{metricwienn}
{\mathit{g\ }_{\mathit{ee}}}&=& - 16(e^{2}\,n^{2} + e^{2}\,\sqrt{
\mathrm{b_n}}\,n^{2} - 2\,e^{2}\,\sqrt{\mathrm{b_n}}\,n + l^{2}\,
\sqrt{\mathrm{b_n}}\,n + l^{2}\,n - 2\,l^{2} - 2\,l^{2}\,\sqrt{
\mathrm{b_n}})\,2^{( - 1/2\,n)} \nonumber \\ &&
(\sqrt{{\displaystyle \frac {n - 2}{n}} }\,l\,\sqrt{1 + \sqrt{
\mathrm{b_n}}})^{n}\,n^{2} \left/ {\vrule
height0.48em width0em depth0.48em} \right. \!  \! ((n - 2)^{2}\,l
^{6}\,(1 + \sqrt{\mathrm{b_n}})^{4}\,(n - 1)\,(n - 4)\,\sqrt{
\mathrm{b_n}}), \nonumber  \\
{\mathit{g\ }_{\mathit{el}}}&=& - 64\, e\,(5\,n^{2}\,e^{4} + e^{4}\,n^{2}
\,\sqrt{\mathrm{b_n}} + 3\,e^{2}\,l^{2}\,\sqrt{\mathrm{b_n}}\,n^{
2} + 5\,n^{2}\,e^{2}\,l^{2} - 6\,e^{2}\,l^{2}\,\sqrt{\mathrm{b_n}
}\,n \nonumber \\ && - 10\,n\,e^{2}\,l^{2}
 + l^{4}\,\sqrt{\mathrm{b_n}}\,n^{2} + l^{4}\,n^{2} - 4\,l
^{4}\,n - 4\,l^{4}\,\sqrt{\mathrm{b_n}}\,n + 4\,l^{4}\,\sqrt{
\mathrm{b_n}} + 4\,l^{4})  \nonumber \\ && 2^{( - 1/2\,n)}
(\sqrt{{\displaystyle \frac {n - 2}{n}} }\,l\,\sqrt{1 + \sqrt{
\mathrm{b_n}}})^{n}\,n^{2} \left/ {\vrule
height0.48em width0em depth0.48em} \right. \!  \! ((n - 2)^{3}\,l
^{9}\,(1 + \sqrt{\mathrm{b_n}})^{6}\,\sqrt{\mathrm{b_n}}\,(n - 1)
), \nonumber  \\
{\mathit{g\ }_{\mathit{ll}}}&=& - 32(30\,l^{6}\,\sqrt{\mathrm{b_n}}
\,n^{2} - 9\,l^{6}\,\sqrt{\mathrm{b_n}}\,n^{3} + l^{6}\,\sqrt{
\mathrm{b_n}}\,n^{4} - 44\,l^{6}\,\sqrt{\mathrm{b_n}}\,n + 108\,n
^{2}\,e^{2}\,l^{4} \nonumber \\ && - 45\,n^{3}\,e^{2}\,l^{4} + 6\,n^{4}\,e^{2}\,l^{4} + 78\,
n^{2}\,e^{4}\,l^{2} - 57\,n^{3}\,e^{4}\,l^{2} + 9\,n^{4}\,e^{4}\,
l^{2} - 84\,n\,e^{2}\,l^{4} \nonumber \\ && + 2\,n^{4}\,e^{6}
+ 24\,l^{6} + l^{6}\,n^{4} - 12\,n^{3}\,e^{6} - 9\,l^{6}
\,n^{3} + 30\,l^{6}\,n^{2} - 44\,l^{6}\,n + 24\,l^{6}\,\sqrt{\mathrm{b_n}}
\nonumber \\ && + 4\,n^{4}\,e^{2}\,l^{4}\,\sqrt{\mathrm{b_n}}
 + 76\,n^{2}\,e^{2}\,l^{4}\,\sqrt{\mathrm{b_n}} - 31\,n^{3
}\,e^{2}\,l^{4}\,\sqrt{\mathrm{b_n}} - 21\,n^{3}\,e^{4}\,l^{2}\,
\sqrt{\mathrm{b_n}} \nonumber \\ && + 30\,n^{2}\,e^{4}\,l^{2}\,\sqrt{\mathrm{b_n}
} - 60\,n\,e^{2}\,l^{4}\,\sqrt{\mathrm{b_n}} + 3\,\sqrt{ \mathrm{b_n}}\,
n^{4}\,e^{4}\,l^{2})\, 2^{( - 1/2\,n)}\, \nonumber \\ && (\sqrt{{\displaystyle \frac {n - 2}{n}} }\,
l\,\sqrt{1 + \sqrt{\mathrm{b_n}}})^{n} \left/ {\vrule height0.48em width0em depth0.48em}
\right. \!  \! ((n - 2)^{3} l^{10}\,(1 + \sqrt{\mathrm{b_n}})^{5}\,(n - 1)\,\sqrt{\mathrm{b_n }}).
\end{eqnarray} \end{scriptsize}

\begin{figure}
\hspace*{0.5cm}
\includegraphics[width=15.0cm,angle=-0]{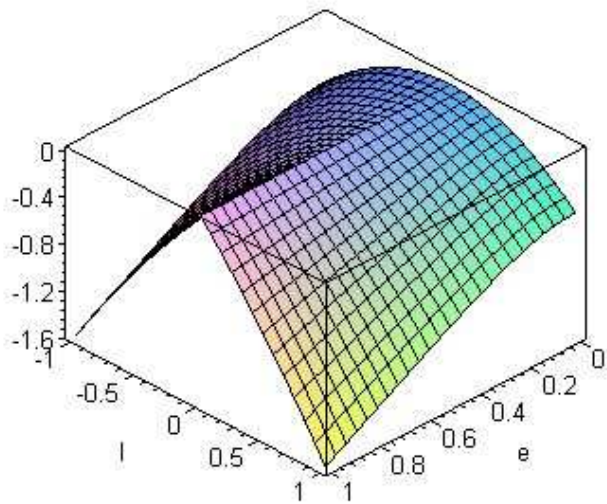}  \vspace*{-5.5cm}
\caption{The $ee$ component of the metric tensor plotted as the function of $\{ e, l \}$,
describing the fluctuations in six dimensional topological Einstein-Yang-Mills black
hole configurations.} \label{ndwee} \vspace*{-0.5cm}
\end{figure}

\begin{figure}
\hspace*{0.5cm}
\includegraphics[width=15.0cm,angle=-0]{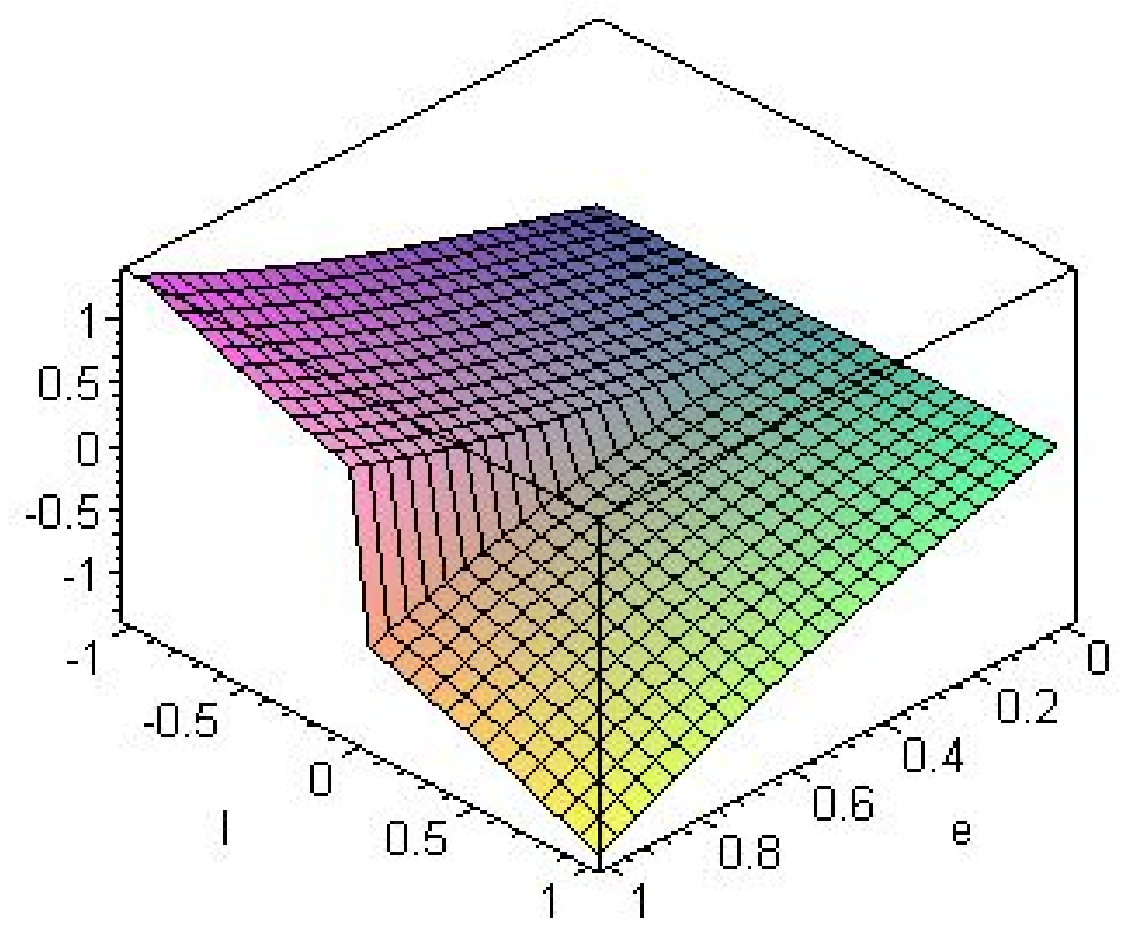}  \vspace*{-5.5cm}
\caption{The $el$ component of the metric tensor plotted as the function of $\{ e, l \}$,
describing the fluctuations in six dimensional topological Einstein-Yang-Mills black
hole configurations.} \label{ndwel} \vspace*{-0.5cm}
\end{figure}

\begin{figure}
\hspace*{0.5cm}
\includegraphics[width=15.0cm,angle=-0]{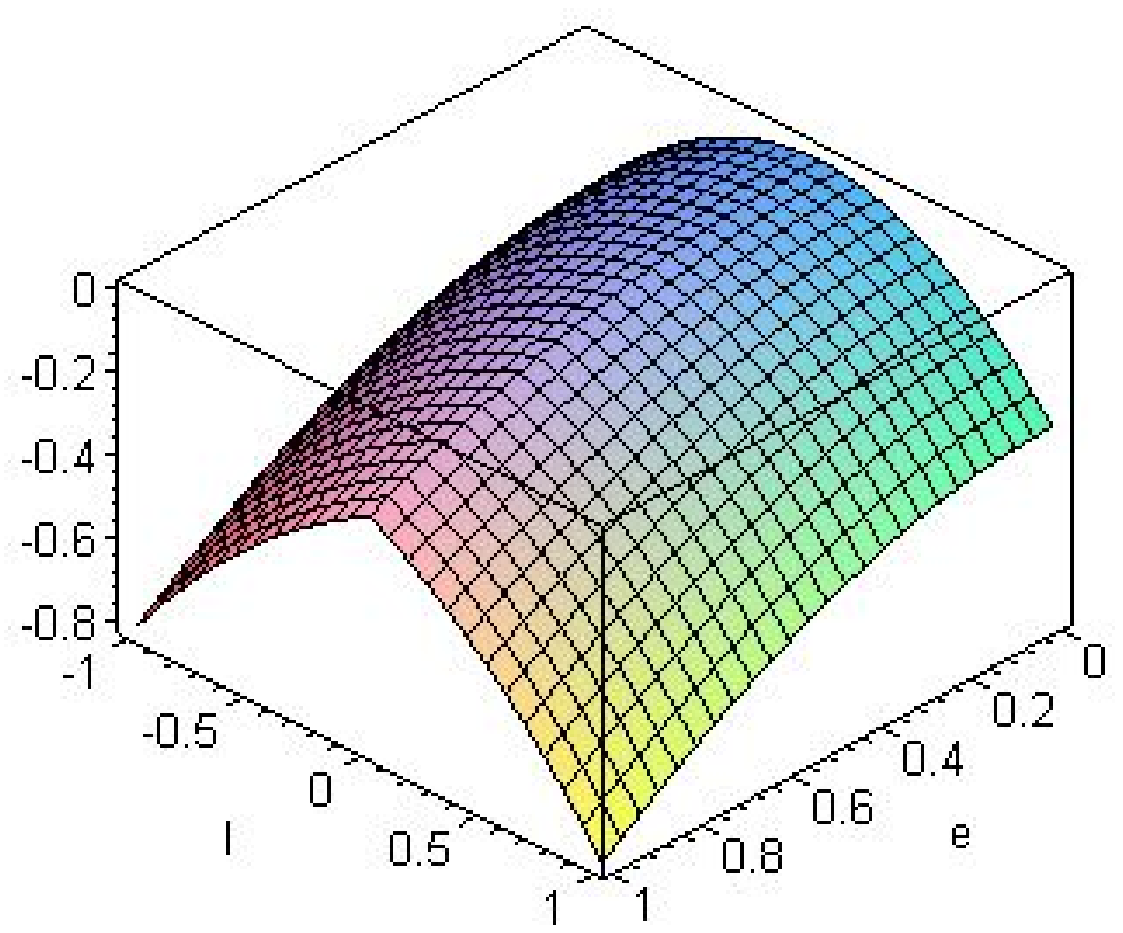}  \vspace*{-5.5cm}
\caption{The $ll$ component of the metric tensor plotted as the function of $\{ e, l \}$,
describing the fluctuations in six dimensional topological Einstein-Yang-Mills black
hole configurations.} \label{ndwll} \vspace*{-0.5cm}
\end{figure}

As mentioned the previous case, we find in this case, in the range of $e \in (0,1)$
and $l \in (-1,1)$ that the fluctuation of the components $\{g_{ee}, g_{ll} \}$ lies
in a negative interval. Namely, the components $g_{ee}$ lies in $(-1.6, 0)$.
While, the components $ g_{ll}$ lies in $(-0.8, 0)$. This shows that the
topological Einstein-Yang-Mills black hole in arbitrary spacetime dimension
$D=1+n$  are locally unstable configurations with respect to the fluctuations
of the $\{e, l\}$. In fact, the range of the growth of $\{g_{ee}, g_{ll} \}$
happens to be in the same negative limit of the amplitude under the flow of the
parameters $\{e, l\}$. Explicitly, from the Figs.(\ref{ndwee}, \ref{ndwll}),
we observe that that the of growth of the $ g_{aa}$ and $g_{bb}$ takes place
in the limit of a large $l$ for all $e$. On the other hand, the component
involving two distinct parameters of black hole has been depicted in
the Figs.(\ref{ndwel}). In this case, we notice that the Figs.(\ref{ndwel})
shows that the $el$-component of the mass fluctuations lies in the interval $(-1.5, 1.5)$.
Thus, for a given ensemble of higher dimensional topological Einstein-Yang-Mills black holes,
the components of the metric tensor $\{g_{ij}\ |\ i, j= e, l \}$ indicate that the
fluctuations involving the vacuum parameters $\{e, l\}$ have relatively
meager positive numerical values and thus they are prone to yield a
statistically unstable ensembl at this order of the ADM mass.

\begin{figure}
\hspace*{0.5cm}
\includegraphics[width=15.0cm,angle=-0] {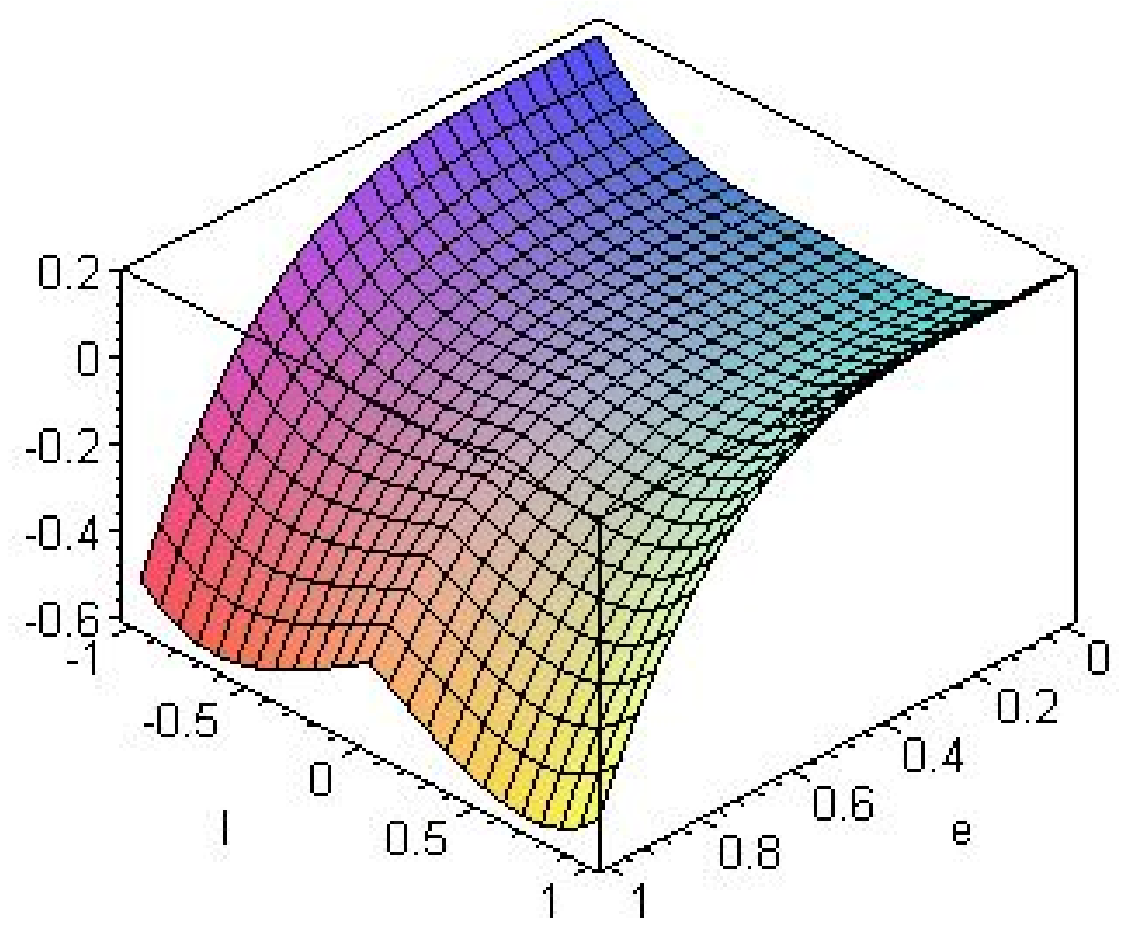}  \vspace*{-5.5cm}
\caption{The determinant of the metric tensor plotted as the function of $\{ e, l \}$,
describing the fluctuations in six dimensional topological Einstein-Yang-Mills black
hole configurations.} \label{ndwdetg} \vspace*{-0.5cm}
\end{figure}

To discuss the thermodynamic properties of arbitrary topological
Einstein-Yang-Mills black hole in $\mathit{(1+n)}$ spacetime dimensions,
we introduce a new scaling function
\begin{scriptsize}
\begin{eqnarray}
\mathrm{c_n} := l^{2}\,n - 2\,l^{2} + 4\,n\,e^{2}
= (n - 2)\,l^{2}\,\mathrm{b_n}.
\end{eqnarray}
\end{scriptsize}

With this convension, we obtain the following Christofel tensors

\begin{scriptsize} \begin{eqnarray}
{\Gamma _{\mathit{eee}}}&=& - \frac{162^{( - 1/2\,n)}\,(\sqrt{
{\displaystyle \frac {n - 2}{n}} }\,l\,\sqrt{1 + \sqrt{
{\displaystyle \frac {\mathrm{c_n}}{(n - 2)\,l^{2}}} }})^{n}\,n^{3}\,e}
{((n - 2)^{2}\,l^{8}\,(1 + \sqrt{{\displaystyle \frac {\mathrm{c_n}}{(n
- 2)\,l^{2}}} })^{5}\,(n - 1) \sqrt{{\displaystyle \frac{\mathrm{c_n}}{(n - 2)\,l^{2}}} }\,\mathrm{c_n})}
({\Gamma _{\mathit{eee}}}^{(1)}+\,\sqrt{{\displaystyle \frac {\mathrm{c_n}}{(n - 2)\,l^{2}}} }\,
{\Gamma _{\mathit{eee}}}^{(2)}), \nonumber  \\
{\Gamma _{\mathit{eel}}}&=& - \frac{1282^{( - 1/2\,n)}\,(\sqrt{
{\displaystyle \frac {n - 2}{n}} }\,l\,\sqrt{1 + \sqrt{{\displaystyle
\frac{\mathrm{c_n}}{(n - 2)\,l^{2}}} }})^{n}\,n^{2}}
{((n - 2)^{3}\,l^{11}\,(1 + \sqrt{{\displaystyle
\frac{\mathrm{c_n}}{(n - 2)\,l^{2}}} })^{8}\,
\sqrt{{\displaystyle \frac {\mathrm{c_n}}{(n - 2)
\,l^{2}}} }\,(n - 1) \mathrm{c_n})}({\Gamma _{\mathit{eel}}}^{(1)}
+\,\sqrt{{\displaystyle \frac {\mathrm{c_n}}{(n - 2)\,l^{2}}} }\,
{\Gamma _{\mathit{eel}}}^{(2)} ), \nonumber  \\
{\Gamma _{\mathit{ele}}}&=& - \frac{1282^{( - 1/2\,n)}\,(\sqrt{
{\displaystyle \frac {n - 2}{n}} }\,l\,\sqrt{1 + \sqrt{
{\displaystyle \frac {\mathrm{c_n}}{(n - 2)\,l^{2}}} }})^{n}\,n^{2}}
{((n - 2)^{3}\,l^{11}\,(1 + \sqrt{{\displaystyle
\frac{\mathrm{c_n}}{(n - 2)\,l^{2}}} })^{8}\,
\sqrt{{\displaystyle \frac {\mathrm{c_n}}{(n - 2)\,l^{2}}} }\,
(n - 1) \mathrm{c_n})}({\Gamma _{\mathit{ele}}}^{(1)}+\,
\sqrt{{\displaystyle \frac {\mathrm{c_n}}{(n - 2)\,l^{2}}} }\,
{\Gamma _{\mathit{ele}}}^{(2)}), \nonumber  \\
{\Gamma _{\mathit{ell}}}&=& -\frac{ 64(\sqrt{{\displaystyle
\frac{n - 2}{n}} }\,l\,\sqrt{1 + \sqrt{{\displaystyle \frac{
\mathrm{c_n}}{(n - 2)\,l^{2}}} }})^{n}\, e\,2^{( - 1/2\,n)}\,n^{2}}
{( (n - 2)^{4}\,l^{12}\,(1 + \sqrt{{\displaystyle
\frac{\mathrm{c_n}}{(n - 2)\,l^{2}}} })^{7}\,(n - 1)\,\sqrt{{\displaystyle
\frac{\mathrm{c_n}}{(n - 2)\,l^{2}}} }\,\mathrm{c_n})}
({\Gamma _{\mathit{ell}}}^{(1)}+ \sqrt{{\displaystyle \frac {\mathrm{c_n}}{(n - 2)\,l^{2}}} }
\,{\Gamma _{\mathit{ell}}}^{(2)}), \nonumber  \\
{\Gamma _{\mathit{lle}}}&=& - \frac{128\,e\,2^{( - 1/2\,n)}\,(\sqrt{{\displaystyle
\frac {n - 2}{n}} }\,l\,\sqrt{1 + \sqrt{{\displaystyle \frac {
\mathrm{c_n}}{(n - 2)\,l^{2}}} }})^{n}\,n^{2} }
{( (n - 2)^{4}\, l^{12}\,(1 + \sqrt{{\displaystyle \frac {\mathrm{c_n
}}{(n - 2)\,l^{2}}} })^{8}\,(n - 1)\,\sqrt{{\displaystyle \frac {
\mathrm{c_n}}{(n - 2)\,l^{2}}} }\,\mathrm{c_n})}
({\Gamma _{\mathit{lle}}}^{(1)}+ \,\sqrt{{\displaystyle \frac {\mathrm{c_n}}{(n - 2)}}}\,
{\Gamma _{\mathit{lle}}}^{(2)}), \nonumber  \\
{\Gamma _{\mathit{lll}}}&=& - \frac{64\,
(\sqrt{{\displaystyle \frac {n - 2}{n}} }\,
l\,\sqrt{1 + \sqrt{{\displaystyle \frac {\mathrm{c_n}}{(n - 2)\,l^{2}}} }})^{n}}
{((n - 2)^{4}\,l^{13}\,(1 + \sqrt{{\displaystyle \frac {\mathrm{c_n
}}{(n - 2)\,l^{2}}} })^{7}\,\sqrt{{\displaystyle \frac {\mathrm{
c_n}}{(n - 2)\,l^{2}}} }\,\mathrm{c_n}\,(n - 1))}({\Gamma _{\mathit{lll}}}^{(1)}+
\sqrt{{\displaystyle \frac {\mathrm{c_n}}{(n - 2)\,l^{2}}} }\, {\Gamma _{\mathit{lll}}}^{(2)}),
\end{eqnarray} \end{scriptsize}

where the factors $\mathit{\{ {\Gamma _{\mathit{ijk}}}^{(1)},
{\Gamma _{\mathit{ijk}}}^{(2)}| i,j,k \in \{e, l\} \}}$
are given as per the followings

\begin{scriptsize} \begin{eqnarray}
{\Gamma _{\mathit{eee}}}^{(1)}&=&
4\,n^{2}\,e^{4}
+ n^{2}\,e^{2}\,l^{2}
+ 8\,n\,e^{2}\,l^{2}
+ 3\,l^{4}\,n
- 6\,l^{4},  \nonumber \\
{\Gamma _{\mathit{eee}}}^{(2)}&=&
e^{2}\,l^{2}\,n^{2}
+ 2\,e^{2}\,l^{2}\,n
+ 3\,l^{4}\,n
- 6\,l^{4},  \nonumber \\
{\Gamma _{\mathit{eel}}}^{(1)}&=&
e^{2}\,l^{6}\,n^{3}
+ e^{2}\,n^{4}\,l^{6}
+ 20\,e^{2}\,l^{6}\,n
- 16\,e^{2}\,l^{6}\,n^{2}  \nonumber \\ &&
+ 4\,e^{4}\,l^{4}\,n^{2}
- 18\,e^{4}\,l^{4}\,n^{3}
+ 8\,e^{4}\,n^{4}\,l^{4}
- 31\,e^{6}\,l^{2}\,n^{3}  \nonumber \\ &&
+ 19\,e^{6}\,n^{4}\,l^{2}
- 8\,l^{8} + n^{3}\,l^{8}
+ 12\,n^{4}\,e^{8}
- 6\,n^{2}\,l^{8}  \nonumber \\ &&
+ 12\,l^{8}\,n, \nonumber \\
{\Gamma _{\mathit{eel}}}^{(2)}&=&
2\,n^{4}\,e^{8}
+ 12\,l^{8}\,n
- 6\,l^{8}\,n^{2}
+ l^{8}\,n^{3}  \nonumber \\ &&
+ l^{6}\,n^{4}\,e^{2}
- l^{6}\,n^{3}\,e^{2}
- 17\,l^{2}\,n^{3}\,e^{6}
+ 12\,l^{4}\,n^{2}\,e^{4} \nonumber \\ &&
+ 9 \,l^{2}\,n^{4}\,e^{6}
- 18\,l^{4}\,n^{3}\,e^{4}
- 8\,l^{6}\,n^{2}\,e^{2}
+ 12\,l^{6}\,n\,e^{2}  \nonumber \\ &&
+ 6\,l^{4}\,n^{4}\,e^{4}
- 8\,l^{8},  \nonumber \\
{\Gamma _{\mathit{ele}}}^{(1)}&=&
e^{2}\,l^{6}\,n^{3}
+ e^{2}\,n^{4}\,l^{6}
+ 20\,e^{2}\,l^{6}\,n
- 16\,e^{2}\,l^{6}\,n^{2}  \nonumber \\ &&
+ 4\,e^{4}\,l^{4}\,n^{2}
- 18\,e^{4}\,l^{4}\,n^{3}
+ 8\,e^{4}\,n^{4}\,l^{4}
- 31\,e^{6}\,l^{2}\, n^{3}  \nonumber \\ &&
+ 19\,e^{6}\,n^{4}\,l^{2}
- 8\,l^{8}
+ n^{3}\,l^{8}
+ 12\,n^{4}\,e^{8} \nonumber \\ &&
- 6\,n^{2}\,l^{8}
+ 12\,l^{8}\,n, \nonumber \\
{\Gamma _{\mathit{ele}}}^{(2)}&=&
2\,n^{4}\,e^{8}
+ 12\,l^{8}\,n
- 6\,l^{8}\,n^{2}
+ l^{8}\,n^{3}  \nonumber \\ &&
+ l^{6}\,n^{4}\,e^{2}
- l^{6}\,n^{3}\,e^{2}
- 17\,l^{2}\,n^{3}\,e^{6}
+ 12\,l^{4}\,n^{2}\,e^{4} \nonumber \\ &&
+ 9\,l^{2}\,n^{4}\,e^{6}
- 18\,l^{4}\,n^{3}\,e^{4}
- 8\,l^{6}\,n^{2}\,e^{2}
+ 12\,l^{6}\,n\,e^{2}  \nonumber \\ &&
+ 6\,l^{4}\,n^{4}\,e^{4}
- 8\,l^{8},  \nonumber \\
{\Gamma _{\mathit{ell}}}^{(1)}&=&
l^{8}\,n^{5}
+ 9\,e^{2}\,n^{5}\,l^{6}
+ 26\,e^{4}\,n^{5}\,l^{4}
+ 25\,e^{6}\,n^{5}\,l^{2}  \nonumber \\ &&
- 13\,l^{8}\,n^{4}
+ 372\,e^{2}\,l^{6}\,n^{3}
- 98\,e^{2}\,n^{4}\,l^{6}
+ 352\,e^{2}\,l^{6}\,n \nonumber \\ &&
- 600\,e^{2}\,l^{6}\,n^{2}
- 508\,e^{4}\,l^{4}\,n^{2}
+ 612\,e^{4}\,l^{4}\,n^{3}
+ 4\,n^{5}\,e^{8} \nonumber \\ &&
- 231\,e^{4}\,n^{4}\,l^{4}
+ 256\,e^{6}\,l^{2}\,n^{3}
- 178\,e^{6}\,n^{4}\,l^{2}
- 80\,l^{8}  \nonumber \\ &&
+ 64\,n^{3}\,l^{8}
- 24\,n^{4}\,e^{8}
- 152\,n^{2}\,l^{8}
+ 176\,l^{8}\,n,  \nonumber \\
{\Gamma _{\mathit{ell}}}^{(2)}&=&
+ l^{8}\,n^{5}
- 13\,l^{8}\,n^{4}
+ 176\,l^{8}\,n
- 152\,l^{8}\,n^{2}  \nonumber \\ &&
+ 64\,l^{8}\,n^{3}
+ 7\,l^{2}\,n^{5}\,e^{6}
+ 14\,l^{4}\,n^{5}\,e^{4}
+ 7\,l^{6}\,n^{5}\,e^{2}  \nonumber \\ &&
- 76\,l^{6}\,n^{4}\,e^{2}
+ 288\,l^{6}\,n^{3}\,e^{2}
+ 76\,l^{2}\,n^{3}\,e^{6}
- 276\,l^{4}\,n^{2}\,e^{4} \nonumber \\ &&
- 52 \,l^{2}\,n^{4}\,e^{6}
+ 332\,l^{4}\,n^{3}\,e^{4}
- 464\,l^{6}\,n^{2}\,e^{2}
+ 272\,l^{6}\,n\,e^{2}  \nonumber \\ &&
- 125\,l^{4}\,n^{4}\,e^{4}
- 80\,l^{8}, \nonumber \\
{\Gamma _{\mathit{lle}}}^{(1)}&=&
l^{8}\,n^{5}
+ 10\,e^{2}\,n^{5}\,l^{6}
+ 34\,e^{4}\,n^{5}\,l^{4}
+ 44\,e^{6}\,n^{5}\,l^{2}  \nonumber \\ &&
- 13\,l^{8}\,n^{4}
+ 414\,e^{2}\,l^{6}\,n^{3}
- 109\,e^{2}\,n^{4}\,l^{6}
+ 392\,e^{2}\,l^{6}\,n  \nonumber \\ &&
- 668\,e^{2}\,l^{6}\,n^{2}
- 664\,e^{4}\,l^{4}\,n^{2}
+ 800\,e^{4}\,l^{4}\,n^{3}
- 302\,e^{4}\,n^{4}\,l^{4}  \nonumber \\ &&
+ 442\,e^{6}\,l^{2}\,n^{3}
- 309\,e^{6}\,n^{4}\,l^{2}
+ 16\,n^{5}\,e^{8}
- 80\,l^{8}  \nonumber \\ &&
+ 64\,n^{3}\,l^{8}
- 88\,n^{4}\,e^{8}
- 152\,n^{2}\,l^{8}
+ 176\,l^{8}\,n,  \nonumber \\
{\Gamma _{\mathit{lle}}}^{(2)}&=&
2\,n^{5}\,e^{8}\,l^{2}
- 13\,l^{8}\,n^{4}
+ l^{8}\,n^{5}
- 12\,n^{4}\,e^{8}  \nonumber \\ &&
+ 176\,l^{8}\,n
- 152\,l^{8}\,n^{2}
+ 64\,l^{8}\,n^{3}
+ 16\,l^{2}\,n^{5}\,e^{6}  \nonumber \\ &&
+ 20\,l^{4}\,n^{5}\,e^{4}
+ 8\,l^{6}\,n^{5}\,e^{2}
- 87\,l^{6}\,n^{4}\,e^{2}
+ 330\,l^{6}\,n^{3}\,e^{2} \nonumber \\ &&
+ 166\,l^{2}\,n^{3}\,e^{6}
- 392\,l^{4}\,n^{2}\,e^{4}
- 115\,l^{2}\,n^{4}\,e^{6}
+ 472\,l^{4}\,n^{3}\,e^{4}  \nonumber \\ &&
- 532\,l^{6}\,n^{2}\,e^{2}
+ 312\,l^{6}\,n\,e^{2}
- 178\,l^{4}\,n^{4}\,e^{4}
- 80\,l^{8}, \nonumber \\
{\Gamma _{\mathit{lll}}}^{(1)}&=&
3360\,n^{5}\,e^{4}\,l^{6}
- 8768\,n^{4}\,e^{6}\,l^{4} + n^{7}\,l^{10}
+ 192\,n^{5}\,e^{10} + 4404\,n^{5}\,e^{6}\,l^{4}  \nonumber \\ &&
+ 41\,n^{7}\,e^{8}\,l^{2}- 384\,l^{10}
- 530\,n^{6}\,e^{8}\,l^{2}
- 6304\,n^{2}\,e^{2}\,l^{8}
+ 11728\,n^{3}\,e^{4}\,l^{6} \nonumber \\ &&
- 924\,n^{6}\,e^{6}\,l^{4}
+ 6544\,n^{3}\,e^{2}\,l^{8}
- 9008\,n^{4}\,e^{4}\,l^{6}
+ 1076\,n^{5}\,e^{2}\,l^{8}  \nonumber \\ &&
+ 43\,n^{7}\,e^{4}\,l^{6}
- 608\,n^{6}\,e^{4}\,l^{6}
- 3568\,n^{4}\,e^{2}\,l^{8}
+ 70\,n^{7}\,e^{6}\,l^{4}   \nonumber \\ &&
- 1552\,l^{10}\,n^{2}
- 17\,l^{10}\,n^{6}
+ 122\,l^{10}\,n^{5}
- 480\,l^{10}\,n^{4}  \nonumber \\ &&
+ 1120\,l^{10}\,n^{3}
+ 1184\,l^{10}\,n
- 56\,n^{6}\,e^{10}
+ 4\,n^{7}\,e^{10} \nonumber \\ &&
+ 2144\,n^{5}\,e^{8}\,l^{2}
+ 11\,n^{7}\,e^{2}\,l^{8}
- 170\, n^{6}\,e^{2}\,l^{8}
+ 2496\,l^{8}\,e^{2}\,n \nonumber \\ &&
+ 6192\,l^{4}\,e^{6}\,n^{3}
- 5952\,l^{6}\,e^{4}\,n^{2}
- 2496\,l^{2}\,n^{4}\,e^{8},  \nonumber \\
{\Gamma _{\mathit{lll}}}^{(2)}&=&
1184\,l^{10}\,n
+ 1120\,l^{10}\,n^{3}
- 384\,l^{10}
- 480\,l^{10}\,n^{4} \nonumber \\ &&
- 1552\,l^{10}\,n^{2}
- 17\,l^{10}\,n^{6}
+ 122\,l^{10}\,n^{5}
+ l^{10}\,n^{7}  \nonumber \\ &&
- 5976\,l^{6}\,n^{4}\,e^{4}
+ 9\,n^{7}\,e^{8}\,l^{2}
+ 556\,n^{5}\,e^{8}\,l^{2}
+ 2060\,l^{4}\,n^{5}\,e^{6}  \nonumber \\ &&
- 4208\,l^{4}\,n^{4}\,e^{6}
- 5312\,l^{8}\,n^{2}\,e^{2}
+ 5488\,l^{8}\,n^{3}\,e^{2}
+ 7872\,l^{6}\,n^{3}\,e^{4} \nonumber \\ &&
+ 2112\,l^{8}\,n\,e^{2}
- 4032\,l^{6}\,n^{2}\,e^{4}
+ 3024\,l^{4}\,n^{3}\,e^{6}
 + 2196\,l^{6}\,n^{5}\,e^{4}  \nonumber \\ &&
+ 30\,n^{7}\,e^{6}\,l^{4}
- 128\,n^{6}\,e^{8}\,l^{2}
- 390\,l^{6}\,n^{6}\,e^{4}
+ 27\,l^{6}\,n^{7}\,e^{4}  \nonumber \\ &&
- 672\,n^{4}\,e^{8}\,l^{2}
+ 892\,l^{8}\,n^{5}\,e^{2}
- 140\,n^{6}\,e^{2}\,l^{8}
+ 9\,n^{7}\,e^{2}\,l^{8}  \nonumber \\ &&
- 416\,n^{6}\,e^{6}\,l^{4}
- 2976\,l^{8}\,n^{4}\,e^{2}. \nonumber \\
\end{eqnarray} \end{scriptsize}

In this case, we find the following determinant of the metric tensor

\begin{scriptsize} \begin{eqnarray}
\mathit{g\ }&=&{\displaystyle \frac {1}{2}} \frac{
((\sqrt{{\displaystyle \frac {n - 2}{2\, n}} }\,l\,\sqrt{1 + \sqrt{{\displaystyle
\frac{\mathrm{c_n}}{(n - 2)\,l^{2}}} }})^{(n - 4)})^{2}}{ (l^{10}\,n^{2}
\,(1 + \sqrt{{\displaystyle \frac {\mathrm{c_n}}{(n - 2)\,l^{2}}
} })^{8} \mathrm{c_n}^{5}\,(n - 4)\,(n - 1)^{2}\,(n - 2)^{2})}
\times {\displaystyle(g_1 + g_2 \sqrt{{\displaystyle \frac {\mathrm{c_n}}{(n - 2)\,l^{2}}} }) }.
\end{eqnarray} \end{scriptsize}

Interestingly, as the functions of $\mathit{\{e,\,l \}}$, it turns out after a simplification
that both the factors $\mathit{g_1(e,\ l)}$ and $\mathit{g_2(e,\ l)}$ can be expressed
as the following finite homogeneous polynomials

\begin{scriptsize} \begin{eqnarray}
g_a:= \sum_{i=1}^{10} g_{ai} \,e^{2i} \,l^{20-2i},\ a=1, 2,
\end{eqnarray} \end{scriptsize}

where both the coefficients $\mathit{\{ g_{1i}, g_{2i} \}}$ can be defined as polynomials
in dimension of the space $\mathit{n\ge 6}$. Explicitly, as the powers of the electric charge
$\mathit{e}$, we find that the coefficients $\mathit{\{ g_{1i}\ | i=1,\dots, 10 \}}$ take the
following degree $\mathit{20}$ expressions

\begin{scriptsize} \begin{eqnarray}
g_{10}&=& - 3145728
+ 16777216\,n
- 40632320\,n^{2}
+ 58982400\,n^{3}
- 57016320\,n^{4} \nonumber \\ &&
+ 38535168\,n^{5}
- 18579456\,n^{6}
 + 6389760\,n^{7}
- 1536000\,n^{8}
+ 245760\,n^{9} \nonumber \\ &&
- 23552\,n^{10}
 + 1024\,n^{11}, \nonumber \\
g_{11}&=&
40894464\,n
- 197656576\,n^{2}
+ 429391872\,n^{3}
- 552075264\,n^{4}
+ 465174528\,n^{5} \nonumber \\ &&
- 268369920\,n^{6}
+ 107347968\,n^{7}
- 29392896\,n^{8}
+ 5271552\,n^{9}
- 559104\,n^{10}  \nonumber \\ &&
+ 26624\,n^{11}, \nonumber \\
g_{12}&=&
- 225705984\,n^{2}
+ 978059264\,n^{3}
- 1880883200\,n^{4}
+ 2106589184\,n^{5} \nonumber \\ &&
- 1514110976\,n^{6}
+ 724140032\,n^{7}
- 230408192\,n^{8}
+ 47022080\,n^{9} \nonumber \\ &&
- 5583872\,n^{10}
+ 293888\,n^{11},  \nonumber \\
g_{13}&=&
680263680\,n^{3}
- 2607677440\,n^{4}
+ 4365025280\,n^{5}
- 4166615040\,n^{6} \nonumber \\ &&
+ 2480128000\,n^{7}
- 942448640\,n^{8}
+ 223211520\,n^{9}
- 30115840\,n^{10} \nonumber \\ &&
+ 1771520\,n^{11},  \nonumber \\
g_{14}&=&
- 1184563200\,n^{4}
+ 3948544000\,n^{5}
- 5626675200\,n^{6}
+ 4442112000\,n^{7} \nonumber \\ &&
- 2097664000\,n^{8}
+ 592281600\,n^{9}
- 92544000\,n^{10}
+ 6169600\,n^{11}, \nonumber \\
g_{15}&=&
1109065728\,n^{5}
- 3142352896\,n^{6}
+ 3696885760\,n^{7}
- 2310553600\,n^{8} \nonumber \\ &&
+ 808693760\,n^{9}
- 150185984\,n^{10}
+ 11552768\,n^{11}, \nonumber \\
g_{16}&=&
- 307888128\,n^{6}
+ 718405632\,n^{7}
- 667090944\,n^{8}
+ 307888128\,n^{9} \nonumber \\ &&
- 70557696\,n^{10}
+ 6414336\,n^{11},  \nonumber \\
g_{17}&=&
- 382205952\,n^{7}
+ 700710912\,n^{8}
- 477757440\,n^{9}
+ 143327232\,n^{10} \nonumber \\ &&
- 15925248\,n^{11},  \nonumber \\
g_{18}&=&
377487360\,n^{8}
- 503316480\,n^{9}
+ 220200960\,n^{10}
- 31457280\,n^{11},  \nonumber \\
g_{19}&=&
- 110100480\,n^{9}
+ 91750400\, n^{10}
- 18350080\,n^{11},  \nonumber \\
g_{110}&=&
6291456\,n^{10}
- 2097152\,n^{11}.
\end{eqnarray} \end{scriptsize}

Similarly, as the powers of the electric charge $\mathit{e}$, it follows
that the second series of factors $\mathit{\{ g_{2i}\ | i=1,\dots, 9 \}}$
reduce to the following degree $\mathit{18}$ polynomials

\begin{scriptsize} \begin{eqnarray}
g_{20}&:=&
- 3145728
+ 16777216\, n
- 40632320\,n^{2}
+ 58982400\,n^{3}
- 57016320\,n^{4}  \nonumber \\ &&
+ 38535168\,n^{5}
- 18579456\,n^{6}
+ 6389760\,n^{7}
- 1536000\,n^{8}
+ 245760\,n^{9}   \nonumber \\ &&
- 23552\,n^{10}
+ 1024\,n^{11},  \nonumber \\
g_{21}&:=&
37748736\,n
- 182452224\,n^{2}
+ 396361728\,n^{3}
- 509607936\,n^{4}
+ 429391872\,n^{5}  \nonumber \\ &&
- 247726080\,n^{6}
+ 99090432\,n^{7}
- 27131904\,n^{8}
+ 4866048\,n^{9}
- 516096\,n^{10}   \nonumber \\ &&
+ 24576\,n^{11},  \nonumber \\
g_{22}&:=&
- 189530112\,n^{2}
+ 821297152\,n^{3}
- 1579417600\,n^{4}
+ 1768947712\,n^{5}  \nonumber \\ &&
- 1271431168\,n^{6}
+ 608075776\,n^{7}
- 193478656\,n^{8}
+ 39485440\,n^{9}  \nonumber \\ &&
- 4688896\,n^{10}
+ 246784\,n^{11}, \nonumber \\
g_{23}&:=&
508035072\,n^{3}
- 1947467776\,n^{4}
+ 3259891712\,n^{5}
- 3111714816\,n^{6}   \nonumber \\ &&
+ 1852211200\,n^{7}
- 703840256\,n^{8}
+ 166699008\,n^{9}
- 22491136\,n^{10}   \nonumber \\ &&
+ 1323008\,n^{11},  \nonumber \\
g_{24}&:=&
- 754384896\,n^{4}
+ 2514616320\,n^{5}
- 3583328256\,n^{6}
+ 2828943360\,n^{7} \nonumber \\ &&
- 1335889920\,n^{8}
+ 377192448\,n^{9}
- 58936320\,n^{10}
+ 3929088\,n^{11},  \nonumber \\
g_{25}&:=&
534773760\,n^{5}
- 1515192320\,n^{6}
+ 1782579200\,n^{7}
- 1114112000\,n^{8} \nonumber \\ &&
+ 389939200\,n^{9}
- 72417280\,n^{10}
+ 5570560\,n^{11},  \nonumber \\
g_{26}&:=&
14155776\,n^{6}
- 33030144\,n^{7}
+ 30670848\,n^{8}
- 14155776\,n^{9} \nonumber \\ &&
+ 3244032\,n^{10}
- 294912\,n^{11}, \nonumber \\
g_{27}&:=&
- 283115520\,n^{7}
+ 519045120\,n^{8}
- 353894400\,n^{9}
+ 106168320\,n^{10}  \nonumber \\ &&
- 11796480\,n^{11}, \nonumber \\
g_{28}&:=&
160432128\,n^{8}
- 213909504\,n^{9}
+ 93585408\,n^{10}
- 13369344\,n^{11}, \nonumber \\
g_{29}&:=&
- 25165824\,n^{9}
+ 20971520\,n^{10}
- 4194304\,n^{11}.
\end{eqnarray} \end{scriptsize}

As the function of $\{e, l\}$, the ensemble stability condition of the
higher dimensional topological Einstein-Yang-Mills black holes follows
from the positivity of the determinant of the metric tensor. In this case,
we find that the determinant of the metric tensor generically tends to a
negative value. For a typical value of $e \in (0,1)$ and $l \in
(-1,1)$, the Fig.(\ref{ndwdetg}) shows that the determinant
of the metric tensor lies in the interval $(-0.6, +0.2)$. Hereby,
for a small $e$, we observe that the determinant of the metric
tensor has an approximate value of $+0.15$. as we increase the
value of the electric charge $e$, in the limit of a large $|l|$,
the determinant of the metric tensor nearly takes a larger positive
value of its amplitude. Thence, for a given $l$, in the limit of a large $e$,
it increases sharply to a larger negative value of order $-0.6$.
When only on of the parameter is allowed to vary, the stability of the
higher dimensional topological Einstein-Yang-Mills black hole
configuration is determined by the positivity of the first
principle minor $p_1:= g_{ee}$, see the Fig.(\ref{ndwee}).
We further find that all the above mentioned qualitative
depictions remain valid for other values of $n$ than the
special case of $n=5$. Thus, the above graphical descriptions
of the principle minors ofers a qualitative notion of the stability
of the an ensemble of topological Einstein-Yang-Mills black holes
in spacetime dimensions $D \ge 5$.

From the Eqn. (\ref{metricwienn}), we see that the Weinhold scalar
curvature of fluctuations vanishes identically. Namely, for all
topological Einstein-Yang-Mills black holes, we find that the
Weinhold scalar curvature vanishes identically and we have

\begin{scriptsize} \begin{eqnarray}
\mathit{R\ }(e, l)=0, \ \forall \ (e,l)\ \in \ \mathcal{M}_2.
\end{eqnarray} \end{scriptsize}

It is surprising to notice that the topological Einstein-Yang-Mills
black holes correspond to a noninteracting statistical system for $D> 5$.
This follows from the vanishing value of the scalar curvature. In
effect, we find it interesting that the scalar curvature vanishes
identically for all values of the black hole parameters $\{e,l\}$.
This shows that the fluctuating higher dimensional topological
Einstein-Yang-Mills black holes are globally stability, namely,
there are no vacuum phase transitions in the underlying ensemble.
In short, the above consideration of the thermodynamic geometry
indicates that the higher dimensional topological Einstein-Yang-Mills black holes
are although noninteracting in the global sense, however they correspond
to a stably vacuum configurations in a specific domain of the parameters.
Namely, when the parameters $\{e, l\}$ are allowed to fluctuate,
there exists certain domain of the $\{e, l\}$ in which there
are positive set of the principle minors. Further, the above observation
follows from the fact that there are non-trivial instabilities at the
local level of the parametric fluctuations of the ensemble.

\section{Conclusion}

The thermodynamic geometric analysis of the topological
Einstein-Yang-Mills black hole configurations is offered
under the fluctuations of the vacuum parameters, namely the
electric charge and the cosmological constant. Such fluctuations
are expected to arise due to non-zero heating effects, chemical reactions
and possible conventional vacuum fluctuation of a field theory configurtaion
containing a background black hole. The intrinsic geometric method is used
to examine the structures of the parametric fluctuations in an ensemble of
arbitrary spacetime dimension topological Einstein-Yang-Mills black holes.
Thus, the stability analysis thus introduced is most generic for the
fluctuations of the parameters that governe the quantum vacuum of the
nonabelian Yang-Mills gauge theory with a background black hole.

The present analysis is well suited for statistical selection of
the stable vacua of the Einstein-Yang-Mills gauge theory. The
thermodynamic geometric procedure is presented for the black holse
carrying a (i) cosmological constant term and (ii) an electric
charge. In this concern, the examination of the thermodynamic
Ruppeiner and Weinhold geometries shows the the entropy and energy
flow properties respectively for an ensemble of
Einstein-Yang-Mills black holes in $D> 5$. The local stability
requires a set of positive heat capacities, while the gloabl
stability of the ensemble requieres the positivity of the
determinant of the metric tensor, as well. These notions
illustrate that the typical instability appears differently for
the case of $D=5$ and $D> 5$ topological Einstein-Yang-Mills black
holes. Subsequently, it turns out that the associated ensembles of
the topological Einstein-Yang-Mills black holes correspond to a
non-interacting system for $D=5$, while it becomes ill-defined for
the case of $D> 5$. From the perspective of the Weinhold geometry,
the first case yields a nonzero scalar curvature, while the second
case leads to a noninteracting statistical configuration. This
follows from the fact that the manifold of parameters is curved,
while it is flat the second case.

In the limit of the electric charge $e \in (0, 10)$ and
cosmological parameter $l \in (-1, +1)$, we find that the
determinant of the Ruppeiner metric tensor of the five dimensional
topological Einstein-Yang-Mills black hole configurations remains
positive, whereas the determinant of the Ruppeiner metric tensor
of the six and higher dimensional topological Einstein-Yang-Mills
black hole configurations vanishes identically for all values of
the electric charge $e$ and the cosmological parameter $l$. This
shows that the five dimensional topological Einstein-Yang-Mills
black hole can be statistically stabilized while the six and
higher dimensional counterparts cannot. Thus, the present
investigation predicts that the topological Einstein-Yang-Mills
black hole systems with the fluctuating $\{e, l\}$ are relatively
more stable for $D=5$ and better-behaved than those in the $D>5$.
In addition, our model is well suited for the robust statisical
examinations. Such black hole configurations are very known now a
days in string theory and M-theory vacuum solutions because of
their existance in the low energy effective field theory
configurations. From the viewpoint vacuum fluctuations, the
parametric fluctuation theory offers a robust model that is very
lucrative. It is worth mentioning that the use of the parametric
geometric principle is rapidly growing in in recent years in the
area of black hole physics.

Based on the notion of the thermodynamic geometries, namely the
definition of Ruppeiner geometry and Weinhold geometry, the
thermodynamic stability analysis remains compatible for
parametrically stable examinations of black holes and their
ensemble properties. The present analysis thus provides the
parametric geometric front to the stability analysis of the
nonabelian topological Yang-Mills black holes and their possible
vacuum fluctuations. The method of the parametric geometry may be
also used, in order to model in a suitable fashion the quantum
part of the background fuctuations and the vacuum disturbances to
the black hole background spacetime configuration. Finally, it is
envisaged that our analysis offers perspective stability grounds,
when applied to the any finite parameter black hole in the
Yang-Mills gauge theory. It is expected further that the present
investigation would be an important factor in an appropriate
examination of the statistical stability creteria of the required
higher derivative corrections and the higher quantum loop
corrections such that the considered black hole ensemble becomes
statistically stable. This consideration can work as the indicator
of fluctuations of the parameters of a given black hole ensemble,
whether yields a statistically (un)stable quantum vacuum.

\section*{Acknowledgement}

Supported in part by the European Research Council grant n.~226455,
\textit{``Supersymmetry, Quantum Gravity and Gauge Fields (SUPERFIELDS)"}.
BNT thanks Prof. V. Ravishankar for support and encouragements towards the present research.
BNT acknowledges the stimulating environments and local hospitality of
(i) \textit{``The Abdus Salam, International Centre for Theoretical Physics Trieste, Italy"},
during the \textit{``Spring School on Superstring Theory and Related Topics (28 March 2011 - 05 April 2011)''};
(ii) \textit{``The Galileo Galilei Institute for Theoretical Physics, Arcetri, Florence''},
during the \textit{``Workshop on Large $N$ Gauge Theories (02 May 2011- 08 May 2011)"};
(iii) \textit{``Les Houches School on Vicious Walkers and Random Martices, (16 May 2011- 27 May 2011)''},
where a part of this research was presented in the Poster Session; and (iv)
\textit{``The Strings 2011, Uppsala Universitet, Sweden (27 June 2011- 02 July 2011)''},
where a part of this research was presented in the Gong-show.
The work of BNT has been supported by a postdoctoral research fellowship of
\textit{``INFN-Laboratori Nazionali di Frascati, Roma, Italy''}.

\end{document}